\documentclass[lettersize,journal]{IEEEtran}
\usepackage{amsmath,amsfonts}
\usepackage{amsthm}
\usepackage{textcomp}
\usepackage{xcolor}
\usepackage{graphicx}
\usepackage{algorithm}
\usepackage[noend]{algorithmic}

\usepackage{multirow}
\usepackage{multicol}
\usepackage{booktabs}
\usepackage{threeparttable}
\usepackage{amsmath}
\usepackage{pifont}
\usepackage{diagbox}
\usepackage{bm}

\newtheorem{theorem}{Theorem}
\newtheorem{definition}{Definition}

\newtheorem {proposition}{Proposition}
\usepackage{amssymb}
\usepackage{tabularray} 
\usepackage{subfigure}
\usepackage{setspace}
\usepackage{epstopdf}
\usepackage{hyperref}

\renewcommand{\algorithmiccomment}[1]{\bgroup\hfill$\triangleright$~#1\egroup}

\usepackage{color}

\begin{document}

\title{Federated Graph Analytics with Differential Privacy}

\author{Shang Liu, Yang Cao, Takao Murakami, Weiran Liu, \\ Seng Pei Liew, Tsubasa Takahashi, Jinfei Liu, Masatoshi Yoshikawa
\thanks{Shang Liu is with Graduate School of Informatics, Kyoto University, Kyoto 606-8501, Japan. Email: liu.shang.33s@st.kyoto-u.ac.jp.
}
\thanks{Yang Cao is with Department of Computer Science, Tokyo Institute of Technology, Tokyo 152-8552, Japan. Email: cao@c.titech.ac.jp.
}
\thanks{Takao Murakami is with Department of Interdisciplinary Statistical Mathematics, Institute of Statistical Mathematics, Tokyo 190-8562, Japan. Email: tmura@ism.ac.jp.
}
\thanks{Weiran Liu is with Alibaba Group, Beijing 100102, China. Email: weiran.lwr@alibaba-inc.com.}
\thanks{Seng Pei Liew and Tsubasa Takahash are with LY Corporation, Tokyo 160–0004, Japan. Email: \{sengpei.liew, tsubasa.takahash\}@lycorp.co.jp.}
\thanks{Jinfei Liu with College of Computer Science and Technology, Zhejiang University, Hangzhou 310027, China. E-mail: jinfeiliu@zju.edu.cn.}
\thanks{Masatoshi Yoshikawa is with Osaka Seikei University, Osaka 533-0007, Japan. Email: yoshikawa-mas@osaka-seikei.ac.jp.}
}

\markboth{JOURNAL OF LATEX CLASS FILES,~Vol.~x, No.~x, AUGUST~2021}%
{Shell \MakeLowercase{\textit{et al.}}: A Sample Article Using IEEEtran.cls for IEEE Journals}


\maketitle

\begin{abstract}
    Collaborative graph analysis across multiple institutions is becoming increasingly popular. 
    Realistic examples include social network analysis across various social platforms, financial transaction analysis across multiple banks, and analyzing the transmission of infectious diseases across multiple hospitals. 
    We define the \textit{federated graph analytics}, a new problem for collaborative graph analytics under differential privacy.
    Although differentially private graph analysis has been
    widely studied, it fails to achieve a good tradeoff between utility and privacy in federated scenarios, due to the \textit{limited view} of local clients and \textit{overlapping information} across multiple subgraphs. 
    Motivated by this, we first propose a \textbf{fe}derated gr\textbf{a}ph analy\textbf{t}ic framework, named $\mathsf{FEAT}$, which enables arbitrary downstream common graph statistics while preserving individual privacy.
    Furthermore, we introduce an optimized framework based on our proposed degree-based partition algorithm, called $\mathsf{FEAT}$+, which improves the overall utility by leveraging the true local subgraphs.
    Finally, extensive experiments demonstrate that our $\mathsf{FEAT}$ and $\mathsf{FEAT+}$ significantly outperform the baseline approach by approximately one and four orders of magnitude, respectively.
\end{abstract}


\section{Introduction}
\label{sec:introduction}

Graph data has become a crucial resource for analyzing big data in a variety of applications such as finance, social networks, and healthcare due to its widespread usage. 
Owing to escalating privacy concerns and regulatory measures like the GDPR, conducting centralized graph analysis has become increasingly challenging.
In this paper, we define the \textit{federated graph analytics} (FGA), a new problem for collaborative graph analysis with the privacy guarantee, which is motivated by the following scenarios:

\vspace{0.5em}
\emph{Example 1.1. Social Network Analytics.}
Various social media platforms, including Facebook, Twitter, and LINE, collaborate to estimate different metrics of a global social network within a particular region. Each platform has its own local subgraph, which is a subset of a ground-truth global social network graph. In a graph, a node represents a user, and an edge represents a friendship between two users. 
As users may use multiple platforms, these clients (i.e., subgraphs) may have \textit{overlapping} edges. 

\emph{Example 1.2. Financial Transaction Analytics.}
\label{example: banks}
Several banks work together to analyze transaction data \cite{cheng2022financial} for financial risk management or macroeconomic analysis over transaction graphs, in which each node represents a bank account owned by a user, and each edge represents a money transaction between two accounts.

\emph{Example 1.3. Disease Transmission Analytics.}
Several medical institutions are collaborating to study the transmission of diseases, such as COVID-19 \cite{dancer2021reducing}, in a particular region. Each hospital is responsible for a subgraph that includes nodes representing patients and edges representing the transmission of the disease between those patients.

\vspace{0.5em}

\begin{figure*}[t]
	\centering  
	\includegraphics[width=0.8\linewidth]{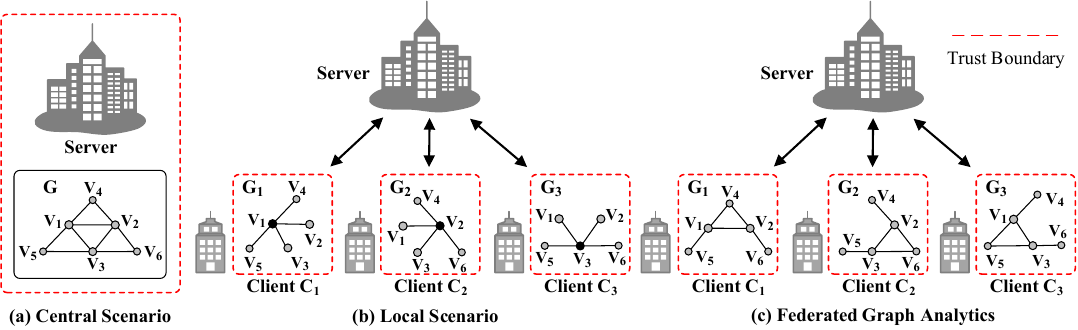}
    \vspace{-0.2cm}
	\caption{Comparisons among central, local and federated scenarios.
 (a) In a central scenario \cite{jian2021publishing,ding2021differentially,day2016publishing,privGraph}, one trusted server owns the entire graph.
 (b) In a local scenario \cite{imola2021locally,ye2020lf,imola2022communication,liu2022collecting}, each client owns one node and its 1-\textit{hop} path information.
 (c) In a federated scenario, each client owns a subgraph that consists of multiple nodes and edges among them. 
 }
\label{fig:three scenarios} 
\end{figure*}

This study is the first to discuss the problem of \textit{federated graph analytics} (FGA) under the \textit{differential privacy} (DP)~\cite{dwork2014algorithmic,li2016differential}.
Different from existing differentially private graph analytic works, such as central models~\cite{jian2021publishing,ding2021differentially,day2016publishing,privGraph,zhang2020community} and local models \cite{imola2021locally,ye2020lf,imola2022communication,liu2022collecting,wei2020asgldp}, FGA considers a more general setting.
In particular, in a central scenario (as shown in Figure \ref{fig:three scenarios}(a)), a trusted server owns the entire global graph that consists of multiple nodes and edges.
Nevertheless, a central server is amenable to privacy issues in practice, such as data leaks and breaches \cite{neto2021developing,gibson2021vulnerability}.
In a local scenario (as shown in Figure \ref{fig:three scenarios}(b)), each client manages an user and her 1-\textit{hop} path information (i.e, neighboring information).
Each client doesn't trust the server and directly perturbs local sensitive data.
In contrast, in federated graph analytics (as shown in Figure \ref{fig:three scenarios}(c)), each client possesses a subgraph consisting of multiple nodes and edges.
Each client does not trust other parties, including the server and other clients.
In fact, local scenarios can be viewed as an extreme case of federated graph analytics when each client contains a subgraph consisting of one user and her 1-hop path.
At this point, $m$ is equal to $n$, where $m$ and $n$ are the number of clients and users, respectively.
Additionally, FGA is similar to cross-silo federated learning~\cite{huang2021personalized,liu2022privacy,zheng2023secure,tang2021incentive} where different silos (or clients) collaboratively train machine learning models without collecting the raw data.
Nevertheless, federated learning focuses on optimization-based questions (i.e., learning models) that differ from graph statistics.

Although differentially private graph analysis has been widely studied \cite{jian2021publishing,ding2021differentially,day2016publishing,privGraph,imola2021locally,ye2020lf,imola2022communication,liu2022collecting,zhang2020community,wei2020asgldp}, this does not apply to FGA due to the following reasons.
On the one hand, the \emph{limited view} of each local client leads to utility issues.
Each client only possesses a portion of the entire graph, making it hard to calculate accurate statistics.
For instance, if a query task $Q$ is to count triangles, each client in Figure \ref{fig:three scenarios}(c) returns the answer $Q_i=1$. 
Although their sum is 3, the true answer is 4. 
This discrepancy happens because the edges of the triangle $\langle v_1,v_2,v_3 \rangle$ come from three different clients. 
Consequently, it is impossible for any individual client to obtain the ground truth answer. 
On the other hand, \emph{overlapping information} among different subgraphs causes privacy issues. 
An edge may exist in multiple subgraphs owned by different clients.
In Figure \ref{fig:three scenarios}(c), for example, the edge $\langle v_1,v_4 \rangle$ appears in both client $C_1$ and client $C_3$. 
Although each client can individually provide sufficient privacy guarantees for  $\langle v_1,v_4 \rangle$, multiple reports of the same information amplify the probability of distinguishing such an edge multiple times, leaking the edge privacy.

In this paper, we propose a \textbf{fe}derated gr\textbf{a}ph analy\textbf{t}ic framework, named $\mathsf{FEAT}$, which enables arbitrary downstream common graph statistics while preserving individual privacy.
The main idea is to let the server privately collect the subgraph information from local clients and then aggregate a noisy global graph for executing targeted query tasks, thereby overcoming the limited view problem.
To avoid collecting the same edge multiple times, $\mathsf{FEAT}$ leverages the private set union (PSU) technique \cite{zhang2023linear,jia2022shuffle,wang2020privacy,dong2017approximating} to aggregate the subgraph information. 
However, existing multi-party private set union protocols do not satisfy DP.
Hence, we design a differentially private set union (DPSU) algorithm, which ensures that the sensitive information is reported only once and the output global graph is protected under DP.

Moreover, we observe that there is still room for improving the accuracy by leveraging true local subgraphs.
To this end, we introduce an improved framework $\mathsf{FEAT}$+ that allows additional communication between the server and clients.
In $\mathsf{FEAT}$+, each client reports the intermediate answer based on its local subgraph and the global graph.
However, a key challenge arises from the possibility of different clients reporting the same edge multiple times, thereby compromising individual privacy. 
To mitigate this risk, we devise a degree-based node partition method to partition entire nodes into multiple disjoint sets. Consequently, the query answer associated with each set is collected only once.

In summary, our contributions in this work are elaborated as follows:

\begin{itemize}
    \item We investigate the federated graph analytics (FGA) under DP for the first time.
    By comparing with previous protocols, we conclude unique challenges in FGA.
    \item We present a generalized federated graph analytic framework with differential privacy ($\mathsf{FEAT}$) based on our proposed DPSU protocol, which supports a wide range of common graph statistics, e.g., subgraph counting.
    \item We introduce an optimized framework ($\mathsf{FEAT}$+) based on our proposed degree-based partition algorithm, which improves the overall utility by leveraging true subgraphs.
    \item We verify the effectiveness of our proposed methods through extensive experiments. $\mathsf{FEAT}$ reduces the error than baseline approach by up to an order of 4. $\mathsf{FEAT}$+ outperforms $\mathsf{FEAT}$ by at least an order of 1.
\end{itemize}

Section \ref{sec:preliminary} introduces the preliminaries. 
Our generalized framework $\mathsf{FEAT}$ and improved framework  $\mathsf{FEAT}$+ are proposed in Section \ref{sec:feat} and Section \ref{sec:feat+}.
Section \ref{sec:experiment} presents experimental results. 
Section \ref{sec:related works} reviews the related work and Section \ref{sec:conclusion} draws a conclusion.

\section{Preliminary}
\label{sec:preliminary}
\subsection{Problem Formulation}
\subsubsection{System Model}
In our work, we consider undirected, unattributed graphs, represented as $G=(V, E)$, where $V=\{v_1,...,v_n\}$ is the set of nodes, and $E\subseteq V \times V$ is the set of edges. 
We study the common graph statistics in cross-silo federated scenario, where there are an untrusted server and $m$ silos, i.e., clients $C=\{C_1,...,C_m\}$.
Each client $C_i$ owns a subgraph $G_i$, which is represented as a $n\times n$ adjacent matrix.
The virtual global graph is the union of all subgraphs, which can be represented as $G=\bigcup_{i=1}^m G_i$.
It is worth noting that each client is mutually independent of others and there may exist overlapping information among different subgraphs, denoted as $G_{i} \cap G_{j} \neq \emptyset$, where $i \neq j$. 
Clients collaboratively support graph queries over their subgraph data while preserving user privacy.
Table \ref{tab:notations} lists the major notations used in this paper.

\subsubsection{Trust Assumption}
Our objective is to create a protocol that enables the server to coordinate graph statistics while ensuring that none of the clients' sensitive information is disclosed. 
Similar to prior works \cite{zheng2023secure,zhang2020batchcrypt,liu2020secure}, we assume that clients are \emph{semi-honest}.
In other words, each client follows the protocol honestly but is curious about the sensitive information on other clients.
The server is untrusted and has no access to individual sensitive information. Furthermore, we presume that any parties beyond the system, such as servers, analysts, or other individuals, are adversaries who are computationally constrained.


\begin{table}[t]
\small
	\caption{Summary of Notations.}
	\label{tab:notations}
	\centering
	\setlength{\tabcolsep}{8mm}{
		\begin{tabular}{l|l}
			\hline
			Notation & Definition \\
			\hline

        $G$ & True global graph \\
        $V$ & Node set \\
        $E$ & Edge set \\
        $G^\prime$ & Noisy global graph \\
        $C$ & A set of all clients \\
        $m$ & Number of clients in $C$ \\
        $C_i$ & The $i$-th client \\
        $G_i$ & Subgraph of client $C_i$\\
        $n$  & Number of nodes in $G$ \\
        $D^\prime$ & Noisy degree sequence\\
        $S_k^\prime$ & Noisy $k$-stars counts\\
        $T^\prime$ & Noisy triangle counts\\
	\hline
	\end{tabular}}
\end{table}

\subsection{Privacy Model}
Like previous works \cite{imola2021locally,ye2020lf,imola2022communication,liu2022collecting,wei2020asgldp}, the private information considered in this study is the edge privacy.
We assume that the server knows the node information, i.e., $V=\{v_1,...,v_n\}$, which makes sense in some real-world applications.
For example, consider that the healthy administration is examining the spread of COVID-19 in a certain region.
It collects the disease transmission paths according to the released census in this area.

Differential privacy (DP)  \cite{dwork2014algorithmic,li2016differential} has become a de-facto standard for preserving individual privacy, which can be formalized in Definition \ref{def:CDP}.
In our work, we use global sensitivity \cite{dwork2014algorithmic} to achieve the DP, defined as Definition \ref{def:global sensitivity}. 
It considers the maximum difference between statistic results on two neighboring graphs. 

\begin{definition}[Differential Privacy \cite{dwork2014algorithmic}]
	\label{def:CDP}
Let $\varepsilon > 0$ be the privacy budget and $n$ be the number of users.
A randomized algorithm $\mathcal{M}$ with domain $\mathbb{D}^n$ satisfies $\varepsilon$-DP, iff for any neighboring datasets $D, D^\prime \in \mathbb{D}^n$ that differ in a single user's data and any subset $S \subseteq Range(\mathcal{M})$, 
	\begin{center}
		$Pr[\mathcal{M}(D) \in S] \leq e^{\epsilon} Pr[\mathcal{M}(D^{\prime}) \in S]$,
	\end{center}  
\end{definition}

\begin{definition}[Global Sensitivity \cite{dwork2014algorithmic}]
	\label{def:global sensitivity}
    For a query function $f$: $D \rightarrow \mathbb{R}$, the global sensitivity is defined by 
       \begin{center}
           $\triangle_{GS}=\mathop{max}\limits_{D\sim D^\prime}|f(D)-f(D^\prime)|$,
       \end{center}
       where $D$ and $D^\prime$ are neighboring databases that differ in a single user's data.
\end{definition}

The Laplace mechanism is one of common techniques to achieve DP. 
The formal definition is as follows:

\begin{definition}[Laplace Mechanism \cite{dwork2006calibrating}]
	\label{def:laplace}
Given any function $f: D \rightarrow R^k$,
let $\triangle f$ be the sensitivity of function $f$.
$M(x)=f(x)+(Y_1,...,Y_k)$ satisfies $(\varepsilon,0)$-differential privacy, where $Y_i$ are $i.i.d$ random variables drawn from Lap($\triangle /\varepsilon$).
\end{definition}

\begin{definition}[Edge LDP \cite{qin2017generating}]
	\label{def:edgeLDP}
For any $i\in[n]$, let $\mathcal{M}_i$ be a randomized algorithm of user $v_i$. 
$\mathcal{M}_i$ satisfies $\varepsilon$-Edge LDP, iff for any two neighboring adjacent bit vectors $A_i$ and $A_i^\prime$ that differ in one edge and any subset $S \subseteq Range(\mathcal{M}_i)$, 
	\begin{center}
		$Pr[\mathcal{M}_i(A_i) \in S] \leq e^{\epsilon} Pr[\mathcal{M}_i(A_i^\prime) \in S]$,
	\end{center} 
where $\varepsilon > 0$ is the privacy budget.
\end{definition}

Existing locally differentially private models, such as $\textit{edge local differentially privacy}$ (Definition \ref{def:edgeLDP}) \cite{imola2021locally,imola2022communication,ye2020lf} is a promising model. However, it fails to provide privacy guarantee in federated scenarios.
While the same edge may exist in multiple different clients, each client only considers its own edge information.
Multiple reports of the same information increase the probability of distinguishing such an edge multiple times, leading to privacy issues. 
To address this challenge, we introduce edge distributed differential privacy to achieve our privacy objectives. 
The formal definition is as follows:

\begin{definition}[Neighboring Graphs]
\label{def:neighboring graphs}
    Two graphs $G$ and $G^\prime$ are neighboring graphs if $G$ and $G^\prime$ differ in one edge.
\end{definition}

\begin{definition}[Edge Distributed Differential Privacy (Edge DDP)]
	\label{def:DDP}
 Let $G=(V,E)$ and $G^\prime=(V,E^\prime)$ be two neighboring global graphs.
Let $C=\{C_1,...,C_m\}$ be the client set.
Let $G_i$ and $G_i^\prime$ ($i\in[1,m]$) be graphs owned by client $C_i$ in $G$ and $G^\prime$, respectively.
A set of randomized mechanisms $\{\mathcal{M}_i, i\in[1,m]\}$ collectively satisfy $\varepsilon$-Edge DDP iff. for any subsets of possible outputs $S_i \subseteq range(\mathcal{M}), i\in[1,m]$, we have the following inequality.
\begin{center}
        $Pr[\mathcal{M}_1(G_1) \in S_1,...,\mathcal{M}_m(G_m) \in S_n]$ \\
    $\leq e^{\epsilon} \cdot Pr[\mathcal{M}_1(G_1^\prime) \in S_1,...,\mathcal{M}_m(G_m^\prime) \in S_m]$.
\end{center}
\end{definition}

Edge DDP guarantees that the server cannot distinguish the presence or absence of any edge based on all reports collected from clients. 
It also guarantees that the information about which client $C_i$ an edge in $E$ belongs to is private, if the edge exists.
For example, in Figure \ref{fig:three scenarios}(c), both the presence of the edge $\langle v_2,v_3\rangle$ in $G$ and the absence of the edge $\langle v_1,v_6\rangle$ in $G$ are private.
Furthermore, no party knows that the edge $\langle v_2,v_3\rangle$ belongs to clients $C_2$ even if the existence of $\langle v_2,v_3\rangle$ has been disclosed.

\subsection{Private Set Union}
Private Set Union (PSU) \cite{zhang2023linear,jia2022shuffle,wang2020privacy,dong2017approximating} is a secure multiparty computation cryptographic technique designed for securely computing the union of private sets held by different parties.
At its core, neither party reveals anything to the counterparty except for the elements in the union.
From a high-level perspective, a typical PSU protocol involves the following steps:
(1) \textit{Set Encoding}: 
each party privately encodes its set into a cryptographic form suitable for secure computation: 
$[\![ X_i ]\!] \leftarrow \mathsf{Enc}(X_i), i\in [1,m]$.
(2) \textit{Union Computation}: 
parties engage in computing the union of their encoded sets without revealing the underlying elements:
$[\![X]\!] \leftarrow \mathsf{Enc}(X_1) \cup \mathsf{Enc}(X_2) ... \mathsf{Enc}(X_m)$.
(3) \textit{Result Decoding}:
once the computation is complete, parties decode the computed union from its cryptographic representation to obtain the set union:
$X \leftarrow \mathsf{Dec}([\![X]\!])$.

\section{A General Framework: $\mathsf{FEAT}$}
\label{sec:feat}
\begin{figure}[t]
	\centering  
	\includegraphics[width=0.85\linewidth]{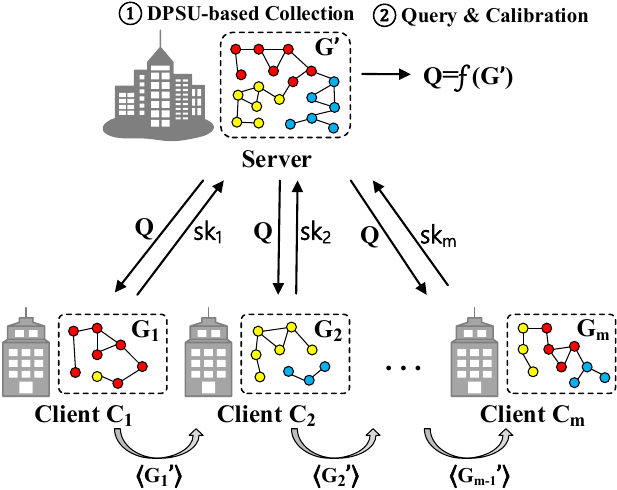}
	\caption{Overview of $\mathsf{FEAT}$.}
\label{fig:FEAT} 
  \vspace{-0.3cm}
\end{figure}

In this section, we introduce our proposed general framework for federated graph analytics with differential privacy, called $\mathsf{FEAT}$.
The main idea is to let the server privately collect subgraphs from local clients and aggregate a noisy global graph, which facilitates common graph statistics.
As shown in Figure \ref{fig:FEAT}, in general, $\mathsf{FEAT}$ enhances the utility by 
(1) reducing the added noise by introducing the crypto technique (i.e., PSU) into differentially private graph statistics;
(2) calibrating the noisy results to suppress the estimation bias.

We first present a baseline approach which revises the prior protocol (e.g., randomized response) in order to satisfy our privacy goal, and discuss its limitations.
Then, we introduce the overview of our proposed general framework $\mathsf{FEAT}$, which substantially improves the baseline approach, and elaborate on its details in subsequent sections.

\subsection{A Baseline Approach}
\label{sec:baseline}

Randomized response (RR) \cite{dwork2014algorithmic} is a common methodology for enhancing local differential privacy.
However, it fails to provide $\varepsilon$-Edge DDP, as the same one edge could be reported by different clients multiple times.
One edge may exist in multiple subgraphs.
In the worst case, an edge is included by all subgraphs $G_i, i\in [1,m]$.
Without loss of generality, assume $G_i$ and $G_i^\prime$ are neighboring graphs and differ in edge $e_1$, we have $Pr[\mathcal{M}_i(G_i) \in S_i] \leq e^{\epsilon} \cdot Pr[\mathcal{M}_i(G_i^\prime) \in S_i]$.
Then, we have
\begin{align*}
    & \frac{Pr[\mathcal{M}_1(G_1) \in S_1,...,\mathcal{M}_m(G_m) \in S_m]}{Pr[\mathcal{M}_1(G_1^\prime) \in S_1,...,\mathcal{M}_m(G_m^\prime) \in S_m]}\\
    = & \frac{Pr[\mathcal{M}_1(G_1) \in S_1]... Pr[\mathcal{M}_m(G_m) \in S_m]}{Pr[\mathcal{M}_1(G_1^\prime) \in S_1]...Pr[\mathcal{M}_m(G_m^\prime) \in S_m]} \\
    \leq & (e^\varepsilon)^m=e^{m\varepsilon}.
\end{align*}

To address this privacy issue, we could consider the following baseline method.
It divides the overall privacy budget into $m$ portions equally.
Then each client randomizes each subgraph using the randomized response with $\varepsilon_l=\varepsilon/m$. 
As a result, the baseline can provide $\varepsilon$-Edge DDP, i.e.,
\begin{align*}
    & \frac{Pr[\mathcal{M}_1(G_1) \in S_1,...,\mathcal{M}_m(G_m) \in S_m]}{Pr[\mathcal{M}_1(G_1^\prime) \in S_1,...,\mathcal{M}_m(G_m^\prime) \in S_m]}\\
    \leq & (e^{\varepsilon_l})^m=e^{m\cdot \frac{\varepsilon}{m}}=e^\varepsilon.
\end{align*}

Algorithm \ref{alg:baseline} shows the details of the baseline approach.
It takes as input the subgraph set $G=\{G_1,...,G_m\}$ and privacy budget $\varepsilon$.
Each graph $G_i$ is represented as a $n\times n$ adjacent matrix $M_i$, where each bit $\in \{0,1\}$.
If there is an edge between two users, the bit will be set as 1; otherwise, it will be set as 0.
After that, each client flips each bit in the upper triangular part of the matrix with the flipping probability $\frac{1}{1+\varepsilon_l}$, where $\varepsilon_l=\varepsilon/m$.
After that, the server collects all noisy edges and aggregates an union as a noisy global graph $G^\prime$.
Finally, the server computes the targeted graph statistics $f(G)^\prime$ based on the noisy global graph $G^\prime$.
This algorithm is denoted as $\mathsf{Baseline}$.

\begin{theorem}
	\label{theorem:privacy baselien}
	$\mathsf{Baseline}$ approach satisfies $\varepsilon$-Edge DDP.
\end{theorem}

\begin{algorithm}[t]
 \small
	\caption{$\mathsf{Baseline}$: Plain Randomized Response} 
	\label{alg:baseline} 
	\begin{algorithmic}[1] 
 
	\REQUIRE  
       \begin{tabular}[t]{l}
         Subgraph set $G=\{G_1,...,G_m\}$, privacy budget $\varepsilon$, \\
         $G_i$ is represented as adjacent matrix $M_i\subseteq\{0,1\}^{n\times n}$\\
      \end{tabular} \\
      
	\ENSURE  
	Noisy global graph $G^\prime$

        Each client $C_i$ perturbs $M_i$\\ 
        \FOR{each bit $b$ in $M_i$}
        \STATE\[
         b\prime=\left \{
         \begin{array}{ll}
            b  & w.p.\ \frac{e^{\varepsilon_l}}{1+e^{\varepsilon_l}} \\
            1-b  & w.p.\ \frac{1}{1+e^{\varepsilon_l}}
         \end{array}
        \right.
       \] \\
        where $\varepsilon_l=\varepsilon/m$
        \ENDFOR
        \STATE Client $C_i$: Send the noisy subgraph $G_i^\prime$ to server
        \STATE Server: $G^\prime \leftarrow \bigcup_{i=1}^{m} G_i^\prime$
	\RETURN $G^\prime$
	\end{algorithmic} 
\end{algorithm}

\textbf{Discussion}.
To provide a strong privacy guarantee, too small privacy budget is allocated to each client.
Although the baseline approach achieves our privacy goal discussed in Section \ref{sec:preliminary}, much redundant noise is added into results.
For example, for $k$-star counting, $\mathsf{Baseline}$ obtains the expected $l_2$ loss errors of $O(\frac{(mn)^{2k-2}}{{\varepsilon}^2})$;
For triangle counting, it attains the expected $l_2$ loss errors of $O(\frac{(mn)^2}{{\varepsilon}^2})$.
Our experiments in Section \ref{sec:experiment} further shows that the baseline approach cannot obtain competitive result accuracy under various cases.

\subsection{Overview of $\mathsf{FEAT}$}
\label{subsec:overview}
To alleviate the limitations of $\mathsf{Baseline}$ method, we propose 
a \textbf{fe}derated gr\textbf{a}ph analy\textbf{t}ic framework, called $\mathsf{FEAT}$, as shown in Figure \ref{fig:FEAT}.
It supports arbitrary downstream common graph statistics while satisfying $\varepsilon$-Edge DDP.

As discussed in the above section, one edge appears in $m$ clients in the extreme case and then the overall privacy budget $\varepsilon$ should be allocated to $m$ clients, i.e., $\varepsilon_l=\varepsilon/m$, which leads to much noise.
In $\mathsf{FEAT}$, we leverage the private set union (PSU) technique \cite{zhang2023linear,jia2022shuffle,wang2020privacy,dong2017approximating} to achieve $\varepsilon_l=\varepsilon$, reducing the noise scale.
PSU allows parties to collaboratively compute the union of multiple sets held by different parties without revealing the individual elements of the sets to other parties.
It guarantees that an element appears in the final union only once.
However, previous PSU protocols cannot be easily applied in our FGA setting.
Most of PSU works \cite{zhang2023linear,jia2022shuffle,dong2017approximating} focus on a two-party setting, which is different from our multi-party scenario.
Although the paper \cite{wang2020privacy} proposes a multi-party protocol, it faces the security and efficiency issue.
Additionally, this multi-party PSU \cite{wang2020privacy} outputs the true union and is unable to provide the differential privacy guarantee. 
To this end, we propose a differentially private set union protocol to compute the union of multiple subgraphs while satisfying $\varepsilon$-Edge DDP.
The detailed analysis will be presented in Section~\ref{subsec:dpsu edge}.

Algorithm \ref{alg:feat} presents the overall protocol of $\mathsf{FEAT}$.
It involves two kinds of entities: clients and a server.
The local clients $C_i$, $i\in[m]$, owns a subgraph that is represented as a $n\times n$ adjacent matrix ($n$ is the number of users).
It takes as input the set of subgraphs $G=\{G_1,...,G_m\}$, privacy budget $\varepsilon$, and the targeted graph query $Q$.
At the beginning, $\mathsf{FEAT}$ recalls a $\mathsf{CollectGraph}()$ function to collect a noisy global graph (step~\ding{172}).
Each client perturbs its subgraph with suitable noise and then encrypts the noisy subgraph.
Next, clients communicate with each other to compute the union of subgraphs.
All clients collaboratively decrypts the union and outputs a noisy global graph (details in Section \ref{subsec:dpsu edge}). 
Once obtaining the noisy global graph, it executes the targeted graph statistics (step~\ding{173}). 
In this paper, we compute the subgraph counts (i.e., $k$-stars, triangles) to verify the effectiveness of our framework. 
Considering the estimation bias, the server further calibrates the noisy results to improve the utility (refer to Section \ref{subsec:feat query} for details).

\begin{algorithm}[t]
  \small
	\caption{Overall Protocol of $\mathsf{FEAT}$.} 
	\label{alg:feat} 
	\begin{algorithmic}[1] 
 
	\REQUIRE  
       \begin{tabular}[t]{l}
         Subgraph set $G=\{G_1,..., G_m\}$, \\
         where $G_i$ represented as $n\times n$ adjacent matrix,\\
         Privacy budget $\varepsilon$,
         Targeted query $Q$\\
      \end{tabular} \\
      
		\ENSURE  
		Query result $Q^\prime$

            \textbf{Step 1: DPSU-based Graph Collection}\\ 
            \STATE $G^\prime \leftarrow \mathsf{CollectGraph}(G, \varepsilon)$    
            \COMMENT{Algorithm \ref{alg:dpsu edge collection}} \\

            \textbf{Step 2: Graph Query and Calibration}\\ 
            \STATE $Q^\prime \leftarrow \mathsf{Query}(G^\prime, Q)$
            \COMMENT{Section \ref{subsec:feat query}} \\ 
            \RETURN $Q^\prime$
	\end{algorithmic} 
\end{algorithm}

\subsection{DPSU-based Graph Collection}
\label{subsec:dpsu edge}

We propose a DPSU protocol based on the state-of-the-art PSU method for computing the global graph under DP. 

$\textbf{PSU Protocol}$.
Suppose that each client $C_{i\in [1,m]}$ owns a set $X_i \subseteq X=\{x_1,...,x_n\}$.
Our goal is to compute the union $X=\bigcup_{i=1}^{m} X_i$.

The SOTA multi-party PSU method~\cite{wang2020privacy} supports computing the union among multiple clients as follows:

\noindent(1) \textit{Initialization}.
Each client $C_i$ creates a flag vector $W_i$.
If $x_{j\in[1,n]} \in X_i$, $W_{i,j}=1$; otherwise, $W_{i,j}=0$.

\noindent(2) \textit{Key Generation}. 
Each client $C_{i\in [1,m]}$ generates a pair of keys $\langle pk_i,sk_i \rangle$, where $pk_i=g^{sk_i}$.
All clients jointly generate the public key: $pk=\prod_{i=1}^m pk_i=g^{sk_1+...+sk_m}$.

\noindent(3) \textit{Encryption}. 
Each client $C_{i\in [1,m]}$ encrypts $W_i$ with the public key $pk$: $\mathsf{Enc}(W_i)=(\mathsf{Enc}(W_{i,1}),...,\mathsf{Enc}(W_{i,n})$.

\noindent(4) \textit{Modification}. 
Each client $C_{i\in [2,m]}$ modifies the flag vector $W_i$ in sequence as follows:
if $x_{j\in[1,n]} \notin X_i$, $\mathsf{Enc}(W_{i,j})=\mathsf{Enc}(W_{i-1,j})$; otherwise, $\mathsf{Enc}(W_{i,j})=\mathsf{Enc}(W_{i,j})$.

\noindent(5) \textit{Decryption}. 
All clients jointly decrypt $\mathsf{Enc}(W_m)$ with secret keys $\{sk_1,...,sk_m\}$: $W_m \leftarrow \mathsf{Dec[Enc}(W_m)]$.
For $j\in[1,n]$, if $W_{m,j}=1$, $X \leftarrow X\cup \{x_j\}$.

However, the above protocol may face the security issue.
It is implemented based on finite field cryptography (FFC)~\cite{diffie2022new} and in fact the prime number $p$ with 512 bits is not secure according to Table 2 in \cite{barker2007sp}.
If this multi-party PSU is implemented in a secure way, the prime number $p$ should be set as 3072 bits.
As a result, it is inefficient for the large-scale graph analytics.
To improve the tradeoff between security and efficiency, we implement the PSU protocol based on the Elliptic Curve Cryptography (ECC) \cite{hankerson2021elliptic}, which is more efficient than finite field cryptography (FFC).
In particular, there are three cryptographic libraries for implementing ECC, namely, Libsodium \cite{libsodium}, OpenSSL \cite{OpenSSL}, and MCL\cite{MCL}.
As shown in Table \ref{tab:ECC}, Libsodium is more suitable for processing large-scale graph data, which motivates us to implement PSU protocol based on Libsodium library.

\begin{table}[t]
\small
	\caption{Performance of ECC with Three Libraries.}
	\centering
    	\label{tab:ECC}
	\setlength{\tabcolsep}{2mm}{
		\begin{tabular}{lrrrrrr}
			\hline
	\diagbox[width=5em]{ECC}{$n$} & $10$  & $10^2$ & $10^3$ & $10^3$ & $10^4$ & $10^5$\\\hline
   Libsodium & 0.0056 & 0.059 & 0.558 & 5.26 & 53.14 & 528.10 \\
   OpenSSL & 0.0098	& 0.060 & 0.549 &	5.49 &	53.76 &	539.52 \\
   MCL & 0.0541 & 0.504 & 4.983 & 49.84 & 498.84 & 4986.43 \\
 \hline
	\end{tabular}}
    \begin{tablenotes}
     \item[1] \quad $n$: data size
   \end{tablenotes}
\end{table}

$\textbf{DPSU Protocol}$.
We then propose a differentially private set union (DPSU) protocol for aggregating a global graph.
One challenge is that the output of the PSU protocol does not satisfy DP.
The individual privacy could be disclosed via the union of subgraphs.
Although some previous works \cite{gopi2020differentially,kim2021differentially,carvalho2022incorporating,qiao2021locally} discuss how to calculate the private set union while satisfying DP, their system models differ from ours and can not be used directly.
One naive solution is that each client $C_i$ perturbs the flag vector $W_i$ using the randomized response (RR) mechanism \cite{warner1965randomized} before encrypting.
Specifically, each client $C_i$ flips each bit in her flag vector $W_i$ with probability $p=\frac{1}{1+e^\varepsilon}$ in the step \textit{Initialization} of PSU protocol, where $\varepsilon$ is the privacy budget.
Subsequently, the differentially private set union can be computed according to step (2)$\sim$(5) in the above PSU protocol.

Although this natural solution can provide the DP guarantee, it leads to much bias in a noisy graph.
In fact, most of real-world graphs tend to be sparse, which means that there are much more 0s than 1s in an adjacent matrix.
After the randomly flipping bits, however, the number of 1s is much larger than that of 0s, making the noisy global graph denser than the original one.
Additionally, the step \textit{Modification} further amplifies the denser problem.
Even if `0' bit is flipped into `1' bit in one of $m$ subgraphs, the according bit in final union will become `1'.
In particular, assume that the number of `1' bits in a true global graph is $t$ and the number of `0' bits is $(n^2-t)$.
After flipping, the number of `1' bits in a noisy global graph becomes $m[(1-p)t+p(n^2-t)]$.
Take the Facebook graph as an example, which has 4,039 nodes (i.e., $n$) and 88,234 edges (i.e., $t$).
Even with a fairly small privacy budget $\varepsilon=0.1$ and the number of clients $m=5$, the expected number of `1' bits will become 3.8$\times 10^7$, increasing at least 439 times than before.

To address the above issue, we propose a novel differentially private set union protocol.
In particular, `0' bits are perturbed only by the first client and `1' bits are randomized by all clients.
The server can then obtain the whole noisy matrix by computing the union of them.
Theoretically, our method can reduce the denser problem by a factor of $m$.
As a result, the utility loss can be alleviated by at least a factor of $O(m^2)$.
For instance, for $k$-star counting, $\mathsf{FEAT}$ achieves the expected $l_2$ loss errors of $O(\frac{n^{2k-2}}{{\varepsilon}^2})$;
For triangle counting, it attains the expected $l_2$ loss errors of $O(\frac{n^2}{{\varepsilon}^2})$.
To further suppress the bias from the randomized response, we calibrate the noisy results during the graph query processing.
The details will be discussed in next Section \ref{subsec:feat query}.

\begin{algorithm}[t]
 \small
\caption{$\mathsf{CollectGraph}$: DPSU-based Graph Collection.} 
	\label{alg:dpsu edge collection} 
	\begin{algorithmic}[1] 
 
	\REQUIRE  
       \begin{tabular}[t]{l}
         Subgraph set $G=\{ G_1,..., G_m\}, G_i=(V_i,E_i),$\\
         Privacy budget $\varepsilon$
      \end{tabular} \\
      
	\ENSURE  
	Noisy global graph $G^\prime$

        \STATE Initialize: An edge domain $\mathbb{E}$ according to $V$,\\
          where $N=|\mathbb{E}|=\frac{n(n-1)}{2}$
        \STATE Server: Initialize $E^\prime=\varnothing$
        \STATE Each client $C_{i\in [1,m]}$ generates a pair of keys $\langle pk_i,sk_i \rangle$, \\
        where $pk_i=g^{sk_i}$
        \STATE All clients jointly generate the public key:\\
        \begin{center}
            $pk=\prod_{i=1}^m pk_i=g^{sk_1+...+sk_m}$
        \end{center}
        \STATE Client $C_1$: Initialize a flag vector $Y$,\\
        \[ Y_{j \in [1,N]}=\left \{
         \begin{array}{ll}
            1,  & \mathbb{E}_{j} \in E_1\\
            0,  & \mathbb{E}_{j} \notin E_1
         \end{array}
        \right.  \]
       \STATE Client $C_1$: Perturb $Y$ with $\mathsf{RR}$,\\
               \[
         Y_{j \in [1,N]}^\prime=\left \{
         \begin{array}{ll}
            Y_j  & w.p. \frac{e^\varepsilon}{1+e^\varepsilon}\\
            1-Y_j  & w.p. \frac{1}{1+e^\varepsilon}\\
         \end{array}
        \right.
       \]
       
      \STATE Client $C_1$: Encrypt $Y^\prime$ with $pk$ to \\
      \begin{center}
          $\mathsf{Enc}(Y^\prime)=[\mathsf{Enc}(Y_1^\prime),...,\mathsf{Enc}(Y_{\widetilde{n}}^\prime)]$
      \end{center} 
      Send $\mathsf{Enc}(Y^\prime)$ to client $C_2$  
      \FOR{each client $C_i, i \in [2,m]$}
      \FOR{each bit $Y_j^\prime, j \in [1,N]$}
      \IF{$\mathbb{E}_{j} \in E_i$}
      \STATE \textbf{if} $\mathsf{RR(1)}==1$ \textbf{then}\quad $\mathsf{Enc}(Y_j^\prime) \leftarrow \mathsf{Enc}(1)$
      \STATE \textbf{else if} $\mathsf{RR(1)}==0$ \textbf{then} \quad $\mathsf{Enc}(Y_j^\prime) \leftarrow \mathsf{Enc}(Y_j^\prime)$
      \ELSIF{$\mathbb{E}_{j} \notin E_i$}
      \STATE \textbf{if} $\mathsf{RR(0)}==1$ \textbf{then}\quad $\mathsf{Enc}(Y_j^\prime) \leftarrow \mathsf{Enc}(1)$
      \STATE \textbf{else if} $\mathsf{RR(0)}==0$ \textbf{then}\quad $\mathsf{Enc}(Y_j^\prime) \leftarrow \mathsf{Enc}(Y_j^\prime)$
      \ENDIF
      
      \ENDFOR
      \ENDFOR
      \STATE All clients jointly decrypt $\mathsf{Enc}(Y^\prime)$ with secret keys:
      \begin{center}
          $Y^\prime \leftarrow \mathsf{Dec[Enc}(Y^\prime)]$
      \end{center}
      Send $Y^\prime$ to server
      \FOR{each bit $Y^\prime, j\in[1,N]$}
      \STATE \textbf{if} $Y_j^\prime=1$ \textbf{then}\quad $E^\prime \leftarrow E^\prime \cup \{\mathbb{E}_j\}$
      \ENDFOR      
      \STATE Server: $G^\prime \leftarrow (V,E^\prime)$
      \RETURN $G^\prime$
	\end{algorithmic} 
\end{algorithm} 

Algorithm \ref{alg:dpsu edge collection} describes the details of our DPSU-based edge collection method.
It takes as input the subgraph set $G$ and the privacy budget $\varepsilon$.
Each client $C_i$ initializes an edge domain $\mathbb{E}$ according to the node information $V$.
In particular, each client $C_i$ first constructs a $n\times n$ adjacent matrix $M_i$, where $n=|V|$, and then transforms the upper triangular part of $M_i$ into a vector $\mathbb{E}$ with the size $N=|\mathbb{E}|=\frac{n(n-1)}{2}$.
For example, $\mathbb{E}_1=e_1=\langle v_1,v_2 \rangle$, $\mathbb{E}_2=e_2=\langle v_1,v_3 \rangle$.
The server initializes an empty set $E$ for collecting the edge information.
Each client $C_i$ generates a pair of keys $\langle ps_i,sk_i \rangle$, and all clients jointly generate the public key $pk$.
Client $C_i$ first initializes a flag vector $Y$ according to the principle in line 5.
If $\mathbb{E}_j$ exists in $E_1$, the according bit will be set as 1; otherwise, it will be set as 0.
After, client $C_1$ perturbs each bit of $Y$ with the flipping probability $\frac{1}{1+e^\varepsilon}$ using the randomized response, and encrypts the noisy $Y^\prime$ with the public key $pk$.
Then, client $C_1$ sends $\mathsf{Enc}(Y^\prime)$ to the client $C_2$.
For each client $C_i$ from $C_2$ to $C_m$, $C_i$ updates $\mathsf{Enc}(Y^\prime)$ according to the principle in lines 10 to 15.
If $\mathbb{E}_j$ is in $E_i$, client $C_i$ will flip `1' with RR and then replace $\mathsf{Enc}(Y_j^\prime)$ with $\mathsf{Enc}(\mathsf{RR}(1))$.
Once the client $C_m$ completes updating $\mathsf{Enc}(Y^\prime)$ and sending it to the server, all clients jointly decrypt $\mathsf{Enc}(Y^\prime)$ and send decrypted $Y^\prime$ to the server.
Finally, the server generates and releases $E$ according to $Y^\prime$.
We call this algorithm by $\mathsf{CollectGraph}$.

\vspace{0.5em}
\emph{Example 3.1.} 
Table \ref{tab:dpsu example} shows how Algorithm \ref{alg:dpsu edge collection} works for computing the union of local edge sets while satisfying the DP.
Specially, clients $C_1, C_2, C_3$ owns private edge sets $E_1=\{e_1,e_2\}, E_2=\{e_2,e_3\}, E_3=\{e_5\}$, and node set is $V=\{v_1,v_2,v_3,v_4\}$.
Thus, $E_1, E_2, E_3 \subseteq \mathbb{E}=\{e_1,e_2,e_3,e_4,e_5,e_6\}$, where $e_1=\langle v_1,v_2\rangle$, $e_2=\langle v_1,v_3\rangle$, $e_3=\langle v_1,v_4\rangle$, $e_4=\langle v_2,v_3\rangle$, $e_5=\langle v_2,v_4\rangle$, $e_6=\langle v_3,v_4\rangle$.
The goal is to compute $E=E_1 \cup E_2 \cup E_3$.
Firstly, client $C_1$ constructs a flag vector $Y$ according to line 5 in Algorithm \ref{alg:dpsu edge collection}.
Then, $C_1$ perturbs each bit of $Y$ and encrypts it to obtain $Y^\prime$.
Secondly, clients $C_2$ and $C_3$ update $Y$ based on the principle of lines 10 to 15 in Algorithm \ref{alg:dpsu edge collection}.
Thirdly, all clients jointly decrypt $\mathsf{Enc}(Y^\prime)$ and send $Y^\prime$ to the server.
Finally, the server can obtain the edge union $E=\{e_1,e_2,e_4,e_5\}$ according to~$Y^\prime$.

\begin{table} [t]
\small
\centering
\caption{An Example of DPSU Protocol.}
\label{tab:dpsu example}
\begin{tblr}{
  width = \linewidth,
  cells = {c},
  hline{1-2,5-7} = {-}{},
}
 $Y$ & $Y_1$ & $Y_2$ & $Y_3$ & $Y_4$ & $Y_5$ & $Y_6$  \\
$C_1$ & $[\![(1)]\!]$ & $[\![(1)]\!]$ & $[\![(0)]\!]$ & $[\![(0)]\!]$  & $[\![(0)]\!]$ & $[\![(0)]\!]$ \\
 
$C_2$ & $[\![(1)]\!]$ & $[\![(1)]\!]$ & $[\![(1)]\!]$& $[\![(0)]\!]$  & $[\![(0)]\!]$ & $[\![(0)]\!]$\\
 
$C_3$ &  $[\![(1)]\!]$ & $[\![(1)]\!]$ & $[\![(1)]\!]$ & $[\![(0)]\!]$  & $[\![(1)]\!]$ & $[\![(0)]\!]$\\
 $Y^\prime$ & 1 & 1 & 0 & 1 & 1 & 0 \\
 $E$ & $e_1$ & $e_2$ & - & $e_4$ & $e_5$ & -\\
\end{tblr}
   \begin{tablenotes}
     \item[1] \quad $[\![x]\!]$: $\mathsf{Enc}(x)$ \quad $(x)$: $\mathsf{RR}(x)$. 
     \item[1] \quad In this example, $Y_3^\prime$ and $Y_4^\prime$ are flipped by RR.
   \end{tablenotes}
     \vspace{-0.4cm}
\end{table}

\begin{theorem}
\label{theorem:edge_collect_dp}
Algorithm \ref{alg:dpsu edge collection} satisfies $\varepsilon$-Edge DDP.
\end{theorem}

\emph{Proof of Theorem \ref{theorem:edge_collect_dp}}. 
Let $G=(V,E)$ and $G^\prime=(V,E^\prime)$ be two neighboring graphs.
Let $G_i$ and $G_i^\prime$ ($i\in[1,m]$) be the neighboring graphs of client $C_i$ with respect to $G$ and $G^\prime$, respectively. 
Let $\{\mathcal{M}_i, i\in[1,m]\}$ be a set of randomized mechanisms with any subsets of possible outputs $S_i \subseteq range(\mathcal{M}_i), i\in[1,m]$.
Given the privacy budget $\varepsilon$, we can easily obtain  $Pr[\mathcal{M}_i(G_i) \in S_i] \leq e^{\varepsilon} Pr[\mathcal{M}_i(G_i^\prime) \in S_i]$.
Due to using the threshold ElGamal encryption, any change of adding or deleting an edge from $m$ clients is only collected by the server once.
Thus, we have 

\begin{align*}
    & \frac{Pr[\mathcal{M}_1(G_1) \in S_1,...,\mathcal{M}_m(G_m) \in S_m]}{Pr[\mathcal{M}_1(G_1^\prime) \in S_1,...,\mathcal{M}_m(G_m^\prime) \in S_m]}\\
    = & \frac{Pr[\mathcal{M}_1(G_1) \in S_1]... Pr[\mathcal{M}_m(G_m) \in S_m]}{Pr[\mathcal{M}_1(G_1^\prime) \in S_1]...Pr[\mathcal{M}_m(G_m^\prime) \in S_m]} \\
    = & e^\varepsilon
\end{align*}

Therefore, Algorithm \ref{alg:dpsu edge collection} satisfies $\varepsilon$-Edge DDP.
Furthermore, by the immunity to post-processing \cite{dwork2014algorithmic}, Algorithm \ref{alg:feat} provides $\varepsilon$-Edge DDP guarantee.
\qed

\subsection{Graph Query Processing}
\label{subsec:feat query}

In this section, we execute two common subgraph counting queries, i.e., $k$-star counting \cite{imola2021locally,imola2022communication,imola2022differentially} and triangle counting \cite{ding2021differentially,imola2021locally,liu2023cargo}, in order to explain how to execute the $\mathsf{Query}$ function of Algorithm~\ref{alg:feat} and how to calibrate the noisy results. 

A $k$-star refers to a subgraph consisting of a central node connecting to $k$ other nodes. 
To count the number of $k$-stars in a given graph, we iterate through each vertex in the graph and compute $\binom{d_i^\prime}{k}$, where $d_i^\prime$ is the noisy degree of node $v_i$.
The $k$-star counts $S$ of the whole graph is equal to the summation of each node's $k$-stars, i.e., $S=\sum_{i=1}^{n} \binom{d_i^\prime}{k}$.
However, a direct computation of node degrees from $G^\prime$ can lead to significant bias induced by the randomized response in Algorithm \ref{alg:dpsu edge collection}.
To mitigate the bias, we leverage the post-processing property of DP \cite{dwork2014algorithmic}, yielding an unbiased estimation $\widetilde{d_i}$ of $d_i$ as Proposition \ref{proposition:feat k-star}.
Hence, we can obtain an unbiased estimation of $k$-stars, i.e., $S=\sum_{i=1}^{n} \binom{\widetilde{d_i}}{k}$.

\begin{proposition}
    \label{proposition:feat k-star}
    Let $G^\prime$ be a noisy global graph.
    Let $d_i^\prime$ be the node degree of $v_i$ in $G^\prime$ and $\widetilde{d_i}$ be an unbiased estimate of $d_i^\prime$.
    Let $n$ be the number of nodes in $G^\prime$.
    Let $p=\frac{1}{1+e^{\varepsilon}}$ be the flipping probability, where $\varepsilon$ is the privacy budget in $\mathsf{FEAT}$.
    We have
    \begin{equation}
         \widetilde{d_i}=\frac{1}{1-2p}(d_i^\prime-np).
    \end{equation}
\end{proposition}

\emph{Proof of Proposition \ref{proposition:debias k-star}}.
The mapping relationship between $d_i^\prime$ and $\widetilde{d_i}$ can be represented as:
\begin{equation}
    d_i^\prime=\widetilde{d_i}(1-p)+(n-\widetilde{d_i})p.
\end{equation}
Then we can prove Proposition \ref{proposition:debias k-star}.
$\qed$

A triangle in a graph refers to a subgraph consisting of three nodes connected by three edges, forming a closed loop.
To count the number of triangles in a graph, one common approach is to iterate each subgraph with three nodes and check if it is a loop.
However, simply counting the triangles in a noisy graph $G^\prime$ introduces a significant bias.
We continue to calibrate the biased triangle counts through the post-processing.
Building upon the insights of \cite{imola2021locally}, we categorize triplets in $G_i$ into four types based on the number of edges they involve: $t_0$, $t_1$, $t_2$, and $t_3$, where $t_j$ ($j \in {0,1,2,3}$) represents the count of triplets in $G_i$ involving $j$ edges (referred to as \textit{$j$-edge}; $3$-edges are identical to triangles). 
Thus, we can compute the unbiased triangle counts $\widetilde{T}$ according to Proposition \ref{proposition:feat triangle}.

\begin{proposition}
    \label{proposition:feat triangle}
    Let $G^\prime$ be a noisy global graph.
    Let $t_0, t_1,t_2,t_3$ be the number of 0-edges, 1-edges, 2-edges, and triangles in $G^\prime$, respectively. 
    Let $\varepsilon$ be the privacy budget used in $\mathsf{FEAT}$.
    We have
    \begin{equation}
        \widetilde{T}=\frac{1}{(e^{\varepsilon}-1)^3}(-t_0+t_1e^{\varepsilon}-t_2e^{2\varepsilon}+t_3e^{3\varepsilon}).
    \end{equation}
\end{proposition}

\emph{Proof of Proposition \ref{proposition:feat triangle}}.
Let $u_0,u_1,u_2,u_3$ be the number of 0-edges, 1-edges, 2-edges, and triangles in $G$, respectively, when we do not flip 1/0 using the randomized response. 
Let $x=e^{\varepsilon}$.
Then we have:
\begin{equation}
(t_0,t_1,t_2,t_3)=(u_0,u_1,u_2,u_3)\bm{A},
\end{equation}
\begin{center}
 $\bm{A}=\frac{1}{(x+1)^2}
\begin{bmatrix}
    x^3 & 3x^2 & 3x & 1 \\
    x^2 & x^3+2x & 2x^2+1 & x \\
    x & 2x^2+1 & x^3+2x & x^2 \\
    1 & 3x & 3x^2 & x^3
\end{bmatrix}$.
\end{center}
$\bm{A}$ is a transition matrix from a type of subgraph (0-edge, 1-edge, 2-edge, and triangle) in an original graph to a type of subgraph in a noisy graph.
Let ($\widetilde{u}_0,\widetilde{u}_1,\widetilde{u}_2$, $\widetilde{u}_3$) be an unbiased estimate of ($u_0, u_1, u_2, u_3$).
Then we obtain:
\begin{equation}
\label{equation:estimation}
(\widetilde{u}_0,\widetilde{u}_1,\widetilde{u}_2,\widetilde{u}_3)=(t_0,t_1,t_2,t_3)\bm{A}^{-1}
\end{equation}
\label{equation:A_i,j}
Let $\bm{A}_{i,j}^{-1}$ be the $(i,j)$-th element of $\bm{A}^{-1}$. 
Then we have:
\begin{equation}
\label{equation:A_1,1}
    \bm{A}_{1,1}^{-1}=\frac{x^3}{(x-1)^3}, \bm{A}_{2,1}^{-1}=-\frac{x^2}{(x-1)^3},
\end{equation}
\begin{equation}
\label{equation:A_2,1}
    \bm{A}_{3,1}^{-1}=\frac{x}{(x-1)^3}, \bm{A}_{4,1}^{-1}=-\frac{1}{(x-1)^3}.
\end{equation}

By combining above equations, Proposition \ref{proposition:feat triangle} is proved. 
\qed

\section{An Improved Framework: $\mathsf{FEAT}$$+$}
\label{sec:feat+}
\begin{figure}[t]
	\centering  
	\includegraphics[width=0.9\linewidth]{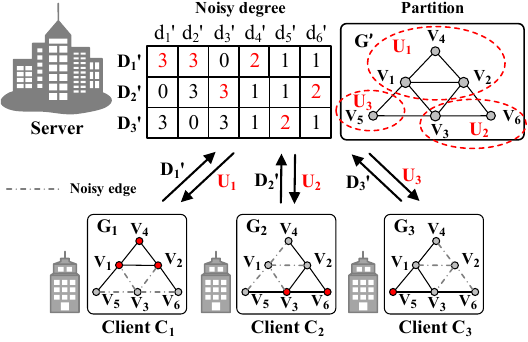}
	\caption{Motivation of $\mathsf{FEAT}$+.
 }
\label{fig:FEAT+} 
\end{figure}

Although $\mathsf{FEAT}$ reduces the problems from the limited view and overlaps, there is still large space for improving the overall utility.
$\mathsf{FEAT}$ calculates common graph statistics based on a noisy global graph collected from local clients and without considering the true local subgraph information.
The graph statistics such as $k$-stars and triangles can be calculated more accurately via an additional round of interaction between the server and clients. 
After the server publishes $G^\prime$, each client $C_i$ can concatenate her local subgraph $G_i$ with $G^\prime$, which removes the limited view issues. 
For example, in Figure \ref{fig:FEAT+}, if the targeted query is to count triangles, client $C_1$ will return $Q_1=4$ instead of $Q_1=1$ since she can see the edges $\langle v_1,v_3\rangle, \langle v_2,v_3\rangle,\langle v_3,v_5\rangle$ and $\langle v_3,v_6\rangle$ via $G^\prime$.
Thus, the final answer will be $(4+4+4)/3=4$ (redundant counts will be removed).

Nevertheless, one edge may be reported by different clients multiple times, which leaks individual privacy (as discussed in Section \ref{sec:introduction}).
For example, in Figure \ref{fig:FEAT+}, the triangle $\langle v_1,v_2,v_3\rangle$ may be collected by the server three times during counting triangles.
In intuition, if one edge is reported by one of $m$ clients only once, the probability of distinguishing such an edge will be reduced.
Hence, the problem is reduced to determining how to assign the same information to one of the clients for reporting.
To do this effectively, we propose a degree-based node partition method that splits nodes into $m$ disjoint node sets so that each node is assigned to a client with the highest node degree. 
In this context, the degree information plays a significant role:
the higher a node degree, the more edges it possesses, and the more likely it is to be involved in $k$-stars or triangles.
For instance, in Figure~\ref{fig:FEAT+}, the degree of node $v_4$ in client $C_1$ ($=2$) is larger than the degrees of clients $C_2$ and $C_3$ ($=1$).
Thus, client $C_1$ can count 2-stars or triangles of $v_4$ more accurately than clients $C_2$ and $C_3$. 
This is an intuition behind our $\mathsf{FEAT}$+. 

\begin{algorithm}[t]
 \small
	\caption{Overall Protocol of $\mathsf{FEAT}$+.} 
	\label{alg:improved overall protocol} 
	\begin{algorithmic}[1] 
 
	\REQUIRE  
       \begin{tabular}[t]{l}
         Subgraph set $G=\{G_1,..., G_m\}$, \\
         Privacy budget $\varepsilon=\varepsilon_1+\varepsilon_2+\varepsilon_3$, sensitivity $\triangle$.\\
      \end{tabular} \\
      
		\ENSURE  
		Query result $Q^\prime$
  
            \textbf{Phase I: Global Graph Collection}\\
            \STATE $G^\prime \leftarrow \mathsf{FEAT}(G,\varepsilon_1)$
            \COMMENT{Algorithm \ref{alg:feat}}\\

            \textbf{Phase II: Local Query Collection} \\
            \STATE $U \leftarrow \mathsf{PartitionNode}(G,\varepsilon_2)$    
            \COMMENT{Step 1: Algorithm \ref{alg:user partition}} \\

            \FOR{each client $C_{i\in[m]}$ in $C$}
            \STATE $Q_i \leftarrow \mathsf{Query}(G^\prime,G_i,U_i)$
             \COMMENT{Step 2: Section \ref{subsec:graph query processing}} \\
            \STATE $Q_i^\prime = Q_i+\mathsf{Lap}(\frac{\triangle}{\varepsilon_3})$
             \COMMENT{Step 3: $\mathsf{Perturbation}$} \\
            \ENDFOR
            \STATE Server: $Q^\prime \leftarrow \sum_{i=1}^m Q_i^\prime$
            \RETURN $Q^\prime$
	\end{algorithmic} 
\end{algorithm}

\subsection{Overview of $\mathsf{FEAT}$+}
Algorithm \ref{alg:improved overall protocol} presents the overall protocol of $\mathsf{FEAT}$+.
It mainly consists of two phases: global graph collection and local query collection.
The global graph collection is used to publish a noisy global graph $G^\prime$, which is finished by $\mathsf{FEAT}$ (Algorithm \ref{alg:feat}).
The local query collection consists of three main steps: degree-based partition (Algortihm \ref{alg:user partition}), graph query processing (Section \ref{subsec:graph query processing}), and perturbation.
To be specific, it takes as input local subgraph set $G=\{G_1,..., G_m\}$, privacy budget $\varepsilon=\varepsilon_1+\varepsilon_2+\varepsilon_3$, and query sensitivity $\triangle$.
$\mathsf{FEAT}$+ first collects a noisy global graph $G^\prime$ via $\mathsf{FEAT}$ using $\varepsilon_1$.
Then, this global graph is published to each client $C_i$ for local query collection.
Next, $\mathsf{FEAT}$+ recalls a $\mathsf{PartitionNode}$ function with privacy budget $\varepsilon_2$, to split the node set $V$ into $m$ disjoint node sets $U$ (Step 1).
These node sets are distributed to the respective clients. 
After that, each client $C_i$ computes the targeted query.
In our paper, we take $k$-star and triangle counting as examples to explain the rationale behind $\mathsf{Query}$ (Step 2).
Since the noisy global graph $G^\prime$ is dense, we carefully design calibration techniques to obtain unbiased estimates of $k$-stars or triangles, as presented in Section \ref{subsec:graph query processing}.
Upon obtaining the query results, each client $C_i$ perturbs the local answers using a suitable noise (Step 3).
Specifically, each client $C_i$ calculates the noisy query $Q_i^\prime = Q_i+\mathsf{Lap}(\frac{\triangle}{\varepsilon_3})$ by adding the Laplacian noise to $Q_i$, where $\triangle$ is the global sensitivity.
Finally, the server computes $Q^\prime \leftarrow \sum_{i=1}^m Q_i^\prime$, which is an unbiased estimate of $Q$ and satisfies DP.

\begin{theorem}
\label{theorem:feat+ privacy}
Algorithm \ref{alg:improved overall protocol} satisfies ($\varepsilon_1+\varepsilon_2+\varepsilon_3$)-Edge DDP.
\end{theorem}

\emph{Proof of Theorem \ref{theorem:feat+ privacy}}.
Algorithm \ref{alg:improved overall protocol} uses three kinds of privacy budgets: 
$\varepsilon_1$ in $\mathsf{FEAT}$,
$\varepsilon_2$ in $\mathsf{PartitionNode}$,
and $\varepsilon_3$ in $\mathsf{Perturbation}$.
By Theorem \ref{theorem:edge_collect_dp}, $\mathsf{FEAT}$ satisfies $\varepsilon_1$-Edge DDP.
In $\mathsf{PartitionNode}$, each client $C_{i\in[m]}$ adds the Laplacian noise $Lap(\frac{m}{\varepsilon_2})$ to the node degree, which satisfies $\varepsilon_2$-Edge DDP.
In $\mathsf{Perturbation}$, each client $C_{i\in[m]}$ adds the Laplacian noise $Lap(\frac{\triangle}{\varepsilon_3})$ to the query result $Q_i$, which satisfies $\varepsilon_3$-Edge DDP.
Following the post-processing property of DP, the aggregated result $Q^\prime$ satisfies $\varepsilon_3$-Edge DDP.
Thus, the entire process of Algorithm~\ref{alg:improved overall protocol} provides ($\varepsilon_1 + \varepsilon_2+\varepsilon_3$)-Edge DDP.
\qed

\subsection{Degree-based Node Partition}
We propose a degree-based user partition technique to split the node set $V$ into multiple disjoint user sets $\{\widetilde{V}_1,...,\widetilde{V}_m\}$.
For each node, the server collects $m$ node degrees $\{d_1^\prime,...,d_m^\prime\}$ from $m$ clients privately.
The node (i.e., user) will be distributed to the client that sends the maximum degree to the server.
Algorithm~\ref{alg:user partition} presents how to partition users into multiple disjoint sets.
It takes as input the true subgraph sets $G=\{V,E\}=\{G_1,..., G_m\}$ and privacy budget $\varepsilon$.
At the outset, the server initializes the empty set $U=\{U_1,...,U_m\}$ for each client, in order to record the partition information.
For each node $v_i$ in the node set $V$, each client $C_{j}$ computes the node degree $d_{i,j}$ based on its true subgraph $G_j$.
Then, client $C_{j}$ perturbs the degree with the Laplace mechanism and sends the noisy degree $d_{i,j}^\prime$ to the server.
After that, the server computes the max noisy degree $d_{i,k}^\prime = \mathsf{max}\{d_{i,1}^\prime,...,d_{i,m}^\prime\}$ for each node $v_i$ and 
adds the user index $i$ to the corresponding user partition set $U_k$.
The final output is the user partition sets $U=\{U_1,...,U_m\}$, where $U_i \cap U_j=\varnothing$ and $\bigcup_{i=1}^{m} U_i=V$. 
This algorithm is denoted by $\mathsf{PartitionNode}$.

\begin{algorithm}[t]
 \small
	\caption{$\mathsf{PartitionNode}$: Degree-based Node Partition.} 
	\label{alg:user partition} 
	\begin{algorithmic}[1] 
 
	\REQUIRE  
       \begin{tabular}[t]{l}
         Subgraph set $G=\{V,E\}=\{G_1,..., G_m\}$, \\
         Privacy budget $\varepsilon_2$\\
      \end{tabular} \\
      
		\ENSURE  
		User partition  $U=\{U_1,...,U_m\}$\\

            \STATE Initialize: $U=\{U_1,...,U_m\}$, where $U_{i\in[m]}=\varnothing$
            \FOR{each node $v_i$ in $V$}
        \STATE Client $C_{j \in [1,m]}$:\begin{tabular}[t]{l} 
             Compute $i$-th node degree $d_{i,j}$ \\
             Perturb             $d_{i,j}^\prime \leftarrow d_{i,j}+\mathsf{Lap}(\frac{m}{\varepsilon_2})$ \\
             Send $d_{i,j}^\prime$ to Server
          \end{tabular}
            \STATE Server: \begin{tabular}[t]{l} 
             Compute $d_{i,k}^\prime \leftarrow \mathsf{max}\{d_{i,1}^\prime,...,d_{i,m}^\prime\}$ \\
             Obtain index $k$ \\
             Update $U_k \leftarrow U_k \cup \{i\}$
          \end{tabular}
            \ENDFOR
            \RETURN $U$
	\end{algorithmic} 
\end{algorithm}

\begin{algorithm}[t]
 \small
	\caption{$k$-Star Counting.} 
	\label{alg:kstar} 
	\begin{algorithmic}[1] 
 
	\REQUIRE  
       \begin{tabular}[t]{l}
       Noisy global graph $G^\prime$, $i$-th subgraph $G_i$\\
       $i$-th user partition $U_i$.\\
      \end{tabular} \\
      
       \ENSURE  
       Number of $k$-stars $S$.

        \STATE Initialize: $d_1=d_2=0$
        \STATE $\overline{G} \leftarrow G^\prime \cup G_i$
        \FOR{each node $v_j$ in $U_i$}
        \FOR{each friend $v_k$ of $v_j$ in $\overline{G}$}
        \STATE \textbf{if} $edge\langle v_j,v_k\rangle$ in $G_i$ \textbf{then} $d_1 \leftarrow d_1+1$
              \COMMENT{True degree in $G_i$} \\
        \STATE \textbf{else}  $d_2 \leftarrow d_2+1$  
              \COMMENT{Noisy degree in $G^\prime$} \\
        \ENDFOR
        \ENDFOR
        \STATE $p \leftarrow \frac{1}{1+e^{\varepsilon_1}}$,
         $\widetilde{d}_2 \leftarrow \frac{1}{1-2p}[d_2-np]$
                      \COMMENT{De-bias} \\
        \STATE $d \leftarrow d_1+\widetilde{d}_2$,
         $S \leftarrow 
         \binom{d}{k}$
        \RETURN $S$
	\end{algorithmic} 
\end{algorithm}

\subsection{Graph Query Processing}
\label{subsec:graph query processing}

In this section, we execute two basic subgraph counting queries, i.e., $k$-star counting \cite{imola2021locally,imola2022communication,imola2022differentially} and triangle counting \cite{ding2021differentially,imola2021locally,liu2023cargo}, in order to explain how to execute the $\mathsf{Query}$ function in line 4 of Algorithm~\ref{alg:improved overall protocol}. 

\subsubsection{$k$-Star Counting}

A $k$-star refers to a subgraph consisting of a central node connecting to $k$ other nodes. 
The key is to compute the total number of edges in a graph, i.e., the summation of node degrees.
From each client $C_i$'s view, each node degree is contributed from a local subgraph $G_i$ and a noisy global graph $G^\prime$.
Thus, the problem is reduced to compute the true node degree $d_1$ and noisy node degree $d_2$.

Algorithm \ref{alg:kstar} shows an instantiation of the $\mathsf{Query}$ function in $k$-star counting.
It takes as input a noisy global graph $G^\prime$, $i$-th subgraph $G_i$, and $i$-th user partition $U_i$.
Client $C_i$ first initializes $d_1=d_2=0$ to record two kinds of degrees.
Then, it obtains a new global graph $\overline{G}$ by computing the union between 
$G^\prime$ and its true subgraph $G_i$ (line 2).
After that, it traverses each node in the user partition set $U_i$ and calculates $d_1$ and $d_2$.
If the edge $\langle v_j,v_k\rangle$ exists in subgraph $G_i$, the true degree will be updated as $d_1 \leftarrow d_1+1$; otherwise, the noisy degree will be updated as $d_2 \leftarrow d_2+1$ (lines 5-6).
However, simply computing the degrees from $G^\prime$ can introduce a significant bias, as the randomized response in Algorithm~\ref{alg:dpsu edge collection} makes a graph dense~\cite{imola2021locally,ye2020lf,qin2017generating}.
By the post-processing property of DP~\cite{dwork2014algorithmic}, we obtain an unbiased estimate $\widetilde{d_2}$ of $d_2$ according to Proposition~\ref{proposition:debias k-star} (line 7).
Upon obtaining the node degree $d=d_1+\widetilde{d_2}$, client $C_i$ can calculate the $k$-stars by 
$\binom{d}{k}$. 
The final output is the number of $k$-stars.

\begin{proposition}
    \label{proposition:debias k-star}
    Let $\overline{G}$ be the union of a noisy global graph $G^\prime$ and local subgraph $G_i$.
    Let $G^C$ be the absolute complement of $G_i$ in $\overline{G}$.
    Let $d_i^\prime$ be the node degree of $v_i$ in $G^C$ and $\widetilde{d_i}$ be an unbiased estimate of $d_i^\prime$.
    Let $n$ be the number of nodes in $G^C$.
    Let $p=\frac{1}{1+e^{\varepsilon_1}}$ be the flipping probability, where $\varepsilon_1$ is the privacy budget in $\mathsf{FEAT}$ (i.e., line 1 of Algorithm~\ref{alg:improved overall protocol}).
    We have
    \begin{equation}
         \widetilde{d_i}=\frac{1}{1-2p}(d_i^\prime-np).
    \end{equation}
\end{proposition}

\emph{Proof of Proposition \ref{proposition:debias k-star}}.
The mapping relationship between $d_i^\prime$ and $\widetilde{d_i}$ can be represented as:
\begin{equation}
    d_i^\prime=\widetilde{d_i}(1-p)+(n-\widetilde{d_i})p.
\end{equation}
Then we can prove Proposition \ref{proposition:debias k-star}.
$\qed$

\subsubsection{Triangle Counting}

\begin{algorithm}[t]
 \small
	\caption{Triangle Counting.} 
	\label{alg:triangle} 
	\begin{algorithmic}[1] 
 
	\REQUIRE  
       \begin{tabular}[t]{l}
       Noisy global graph $G^\prime$, $i$-th subgraph $G_i$\\
       $i$-th user partition $U_i$.\\
      \end{tabular} \\
      
       \ENSURE  
       Number of triangles $T$.

        \STATE Initialize: $T_0=T_1=T_2=T_3=0$
        \STATE $\overline{G} \leftarrow G^\prime \cup G_i$
        \FOR{each node $v_j$ in $U_i$}
        \FOR{each friend $v_k$ of $v_j$ in $\overline{G}$}
        \FOR{each friend $v_l$ of $v_j$ in $\overline{G}$}
        \IF{$j<k<l$}
        \IF{$v_k, v_l$ are friends in $\overline{G}$}
        \STATE Initialize: $\bm{e}=\{ \langle v_j,v_k\rangle,\langle v_j,v_l\rangle,\langle v_k,v_l\rangle\}$
        \STATE \textbf{if} 0 edges of $\bm{e}$ in $G_i$ \textbf{then} $T_0 \leftarrow T_0+1$
         \STATE \textbf{if} 1 edges of $\bm{e}$ in $G_i$ \textbf{then} $T_1 \leftarrow T_1+1$
        \STATE \textbf{if} 2 edges of $\bm{e}$ in $G_i$ \textbf{then} $T_2 \leftarrow T_2+1$
        \STATE \textbf{if} 3 edges of $\bm{e}$ in $G_i$ \textbf{then} $T_3 \leftarrow T_3+1$
        \ENDIF
        \ENDIF
        \ENDFOR
        \ENDFOR
        \ENDFOR
        
        \STATE $(\widetilde{T}_0,\widetilde{T}_1,\widetilde{T}_2) \leftarrow \mathsf{Debias}(T_0,T_1,T_2)$
        \STATE $T \leftarrow \widetilde{T}_0+\widetilde{T}_1+\widetilde{T}_2+T_3$
        \RETURN $T$
	\end{algorithmic} 
\end{algorithm}

Next, we focus on triangle counting. 
This is more challenging because three edges of a triangle can be from the local subgraph $G_i$ or a noisy global graph $G^\prime$.
There are four kinds of triangles according to the number of edges from $G_i$: $T_0,T_1,T_2$, and $T_3$, where $T_j$ ($j \in \{0,1,2,3\}$) is the number of triangles involving $j$ edges from $G_i$ (referred to as \textit{$j$-edge} in $G_i$). 
Since $T_0,T_1$, and $T_2$ involves some noisy edges from $G^\prime$, simply counting the noisy triangles can introduce a bias.
We propose different empirical estimation methods to obtain unbiased counts, as presented in Propositions~\ref{proposition:debias triangle T0}, \ref{proposition:debias triangle T1}, and \ref{proposition:debias triangle T2} at the end of this subsection.

Algorithm \ref{alg:triangle} presents an instantiation of the $\mathsf{Query}$ function in triangle counting.
It takes as input a noisy global graph $G^\prime$, the $i$-th subgraph $G_i$, the $i$-th user partition $U_i$, and the privacy budget $\varepsilon_1$.
Client $C_i$ first obtains a graph $\widetilde{G}$ by merging $G^\prime$ and $G_i$.
Then it calculates four kinds of triangles $(T_0, T_1, T_2, T_3)$, i.e., 0-edge, 1-edge, 2-edge, and 3-edge in $G_i$.
After that, the unbiased estimates $(\widetilde{T}_0, \widetilde{T}_1, \widetilde{T}_2)$ of $(T_0, T_1, T_2)$ are computed based on Proposition \ref{proposition:debias triangle T0}, \ref{proposition:debias triangle T1}, and \ref{proposition:debias triangle T2}.
The final triangle count $T$ can be obtained by aggregating $\widetilde{T}_0, \widetilde{T}_1, \widetilde{T}_2$, and $T_3$.

\begin{proposition}
    \label{proposition:debias triangle T0}
    Let $\overline{G}$ be the union of a noisy global graph $G^\prime$ and local subgraph $G_i$.
    Let $G^C$ be the absolute complement of $G_i$ in $\overline{G}$.
    Let $t_0, t_1,t_2,t_3$ be the number of 0-edges, 1-edges, 2-edges, and triangles in $G^C$, respectively. 
    Let $T_0$ be the number of 0-edges in $G_i$, i.e., $T_0=t_3$.
    Let $\widetilde{T}_0$ be an unbiased estimate of $T_0$.
    Let $\varepsilon_1$ be the privacy budget used in $\mathsf{FEAT}$ (i.e., line 1 of Algorithm~\ref{alg:improved overall protocol}).
    We have
    \begin{equation}
        \widetilde{T}_0=\frac{1}{(e^{\varepsilon_1}-1)^3}(-t_0+t_1e^{\varepsilon_1}-t_2e^{2\varepsilon_1}+T_0e^{3\varepsilon_1}).
    \end{equation}
\end{proposition}

The proof details of Proposition \ref{proposition:debias triangle T0} can refer to Proposition~\ref{proposition:feat triangle}.

\begin{proposition}
    \label{proposition:debias triangle T1}
    Let $\overline{G}$ be the union of a noisy global graph $G^\prime$ and local subgraph $G_i$.
    Let $G^C$ be the absolute complement of $G_i$ in $\overline{G}$.
    Let $t_0, t_1, t_2$ be the number of 0-edges, 1-edges, and 2-edges in $G^C$, respectively. 
    Let $T_1$ be the number of 1-edges in $G_i$, i.e., $T_1=t_2$.
    Let $\widetilde{T}_1$ be an unbiased estimate of $T_1$.
    Let $\varepsilon_1$ be the privacy budget used in $\mathsf{FEAT}$.
    We have
    \begin{equation}
        \widetilde{T}_1=\frac{(e^{\varepsilon_1}+1)[e^{2\varepsilon_1}t_0-e^{\varepsilon_1}(e^{\varepsilon_1}+1)t_1+(2e^{\varepsilon_1}+1)T_1]}{(e^{\varepsilon_1}-1)(e^{2\varepsilon_1}-2e^\varepsilon-1)}.
    \end{equation}
\end{proposition}

\emph{Proof of Proposition \ref{proposition:debias triangle T1}}. 
Consider that one edge of a triangle is from $G_i$ and another two edges are from $G^C$.
Let $u_0,u_1,u_2$ be the number of 0-edges, 1-edges, and 2-edges in $G^C$, respectively, when we do not flip 1/0 using the randomized response. 
Let $x=e^{\varepsilon_1}$.
Then we have:
\begin{equation}
(t_0,t_1,t_2)=(u_0,u_1,u_2)\bm{A}
\end{equation}
\begin{center}
 $\bm{A}=\frac{1}{(x+1)^2}
\begin{bmatrix}
    x^2 & 2x & 1 \\
    x & 1+x & x \\
    1 & 2x & x^2 \\
\end{bmatrix}$   
\end{center}
$\bm{A}$ is a transition matrix from a type of subgraph (0-edge, 1-edge, 2-edge) in an original graph to a type of subgraph in a noisy graph.
Let $\widetilde{u}_0,\widetilde{u}_1,\widetilde{u}_2$ be the unbiased estimation of ($u_0, u_1,u_2$).
Then we obtain:
\begin{equation}
\label{equation:estimation}
(\widetilde{u}_0,\widetilde{u}_1,\widetilde{u}_2)=(t_0,t_1,t_2)\bm{A}^{-1}
\end{equation}
\label{equation:A_i,j}
Let $\bm{A}_{i,j}^{-1}$ be the $(i,j)$-th element of $\bm{A}^{-1}$. 
Then we have:
{
\begin{equation}
\label{equation:A_1,1}
    \bm{A}_{1,1}^{-1}=\frac{(x+1)x^2}{(x-1)(x^2-2x-1)},
\end{equation}
\begin{equation}
\label{equation:A_2,1}
    \bm{A}_{2,1}^{-1}=\frac{(x+1)(-x^2-x)}{(x-1)(x^2-2x-1)},
\end{equation}
\begin{equation}
\label{equation:A_3,1}
    \bm{A}_{3,1}^{-1}=\frac{(x+1)(2x+1)}{(x-1)(x^2-2x-1)},
\end{equation}
}
By combining above equations, Proposition \ref{proposition:debias triangle T1} is proved. 
\qed

\begin{proposition}
    \label{proposition:debias triangle T2}
    Let $G^\prime$ be a noisy global graph and $G_i$ be the local subgraph.
    Let $T_2$ be the number of 2-edges in $G_i$.
    Let $S$ be the number of 2-stars in $G_i$.
    Let $\widetilde{T}_2$ be an unbiased estimate of $T_2$.
    Let $p=\frac{1}{1+e^{\varepsilon_1}}$ be the flipping probability.
    We have
    \begin{equation}
        \widetilde{T}_2=\frac{1}{1-2p}(T_2-Sp).
    \end{equation}
\end{proposition}

\emph{Proof of Proposition \ref{proposition:debias triangle T2}}.
The mapping relationship between $d_i^\prime$ and $\widetilde{d_i}$ can be represented as:
\begin{equation}
    T_2=\widetilde{T}_2(1-p)+(S-\widetilde{T}_2)p.
\end{equation}
Then we can prove Proposition \ref{proposition:debias triangle T2}.
$\qed$

\section{Experimental Evaluation}
\label{sec:experiment}
In this section, we evaluate our $\mathsf{FEAT}$ and $\mathsf{FEAT}$+ along two dimensions: utility and running time.
To simulate the federated scenario, we split a graph randomly into multiple local subgraphs by controlling two key parameters: \emph{sampling rate} $\rho$ and \emph{overlapping rate} $\sigma$.
Then, we conducted experiments to answer the following questions:
\begin{itemize}
  \item \textbf{Q1}: How do our general $\mathsf{FEAT}$ and improved $\mathsf{FEAT}$+ compare with baseline approaches (denoted by $\mathsf{Baseline}$) in terms of the utility-privacy trade-off?
  \item \textbf{Q2}: How do the sampling rate $\rho$ and overlapping rate $\sigma$ affect accuracy?
  \item \textbf{Q3}: How much do our $\mathsf{FEAT}$ and $\mathsf{FEAT}$+ compare with $\mathsf{Baseline}$ in terms of the running time?
\end{itemize}

\begin{figure*}[t]
 \begin{center}
     \begin{minipage}[t]{0.49\linewidth}
   \centering
\subfigure[Facebook]{
		\begin{minipage}[t]{0.49\linewidth}
			\centering
			\includegraphics[width=\linewidth]{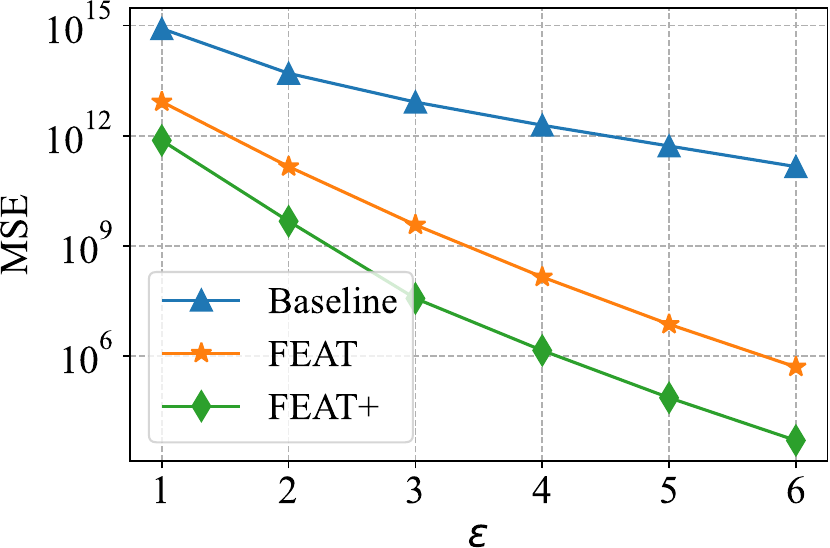}
		\end{minipage}
	}%
\subfigure[Wiki]{
		\begin{minipage}[t]{0.49\linewidth}
			\centering
			\includegraphics[width=\linewidth]{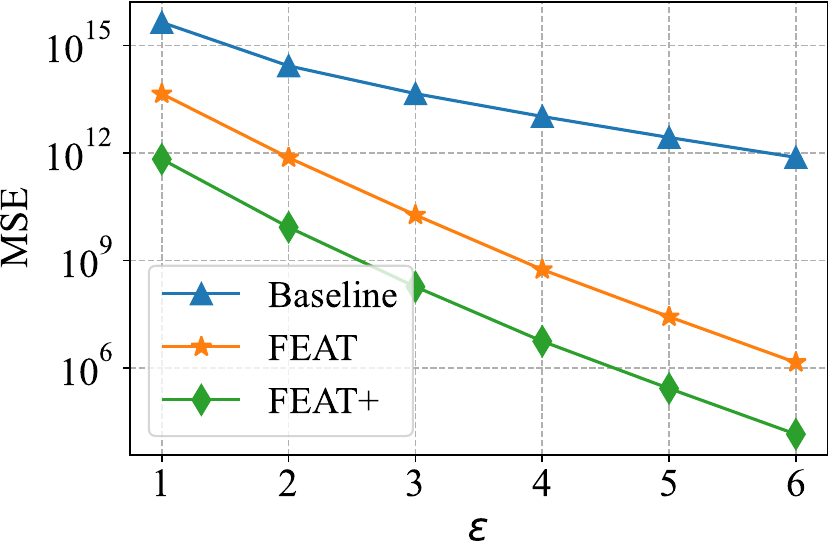}
		\end{minipage}%
    \label{fig:2star mse wiki}
	}
      \vspace{-0.4cm}
       \caption{The MSE in 2-star counting.}
       \label{fig:2star mse}
\end{minipage}
 \begin{minipage}[t]{0.49\linewidth}
   \centering
\subfigure[Facebook]{
		\begin{minipage}[t]{0.49\linewidth}
			\centering
			\includegraphics[width=\linewidth]{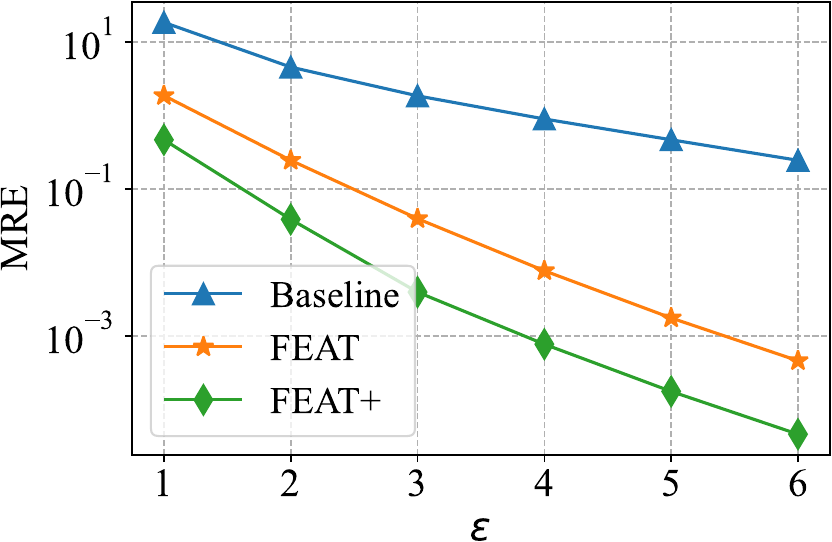}
		\end{minipage}
  \label{fig:2star mre facebook}
	}%
\subfigure[Wiki]{
		\begin{minipage}[t]{0.49\linewidth}
			\centering
			\includegraphics[width=\linewidth]{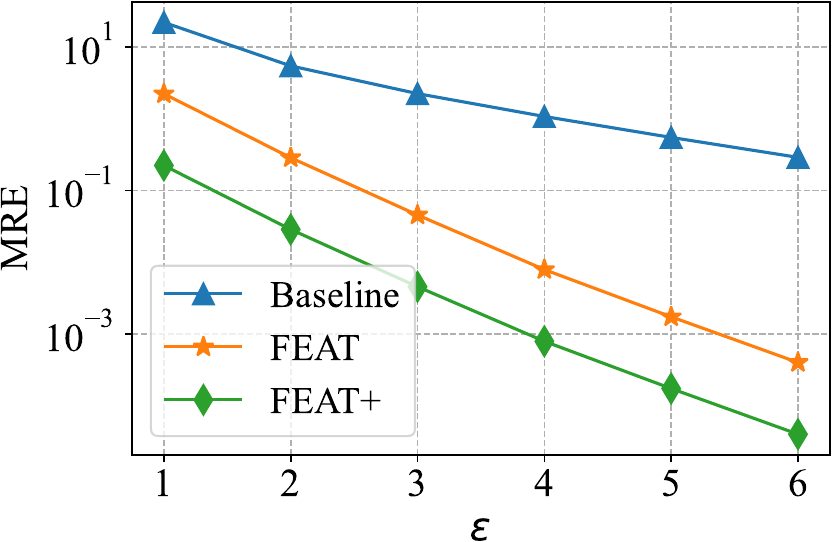}
		\end{minipage}%
    \label{fig:2star mre wiki}
	}
       \vspace{-0.4cm}
       \caption{The MRE in 2-star counting.}
        \label{fig:2star mre}
\end{minipage}
   \end{center}
        \vspace{-0.3cm}
\end{figure*}

\begin{figure*}[t]
 \begin{center}
     \begin{minipage}[t]{0.49\linewidth}
   \centering
\subfigure[Facebook]{
		\begin{minipage}[t]{0.49\linewidth}
			\centering
			\includegraphics[width=\linewidth]{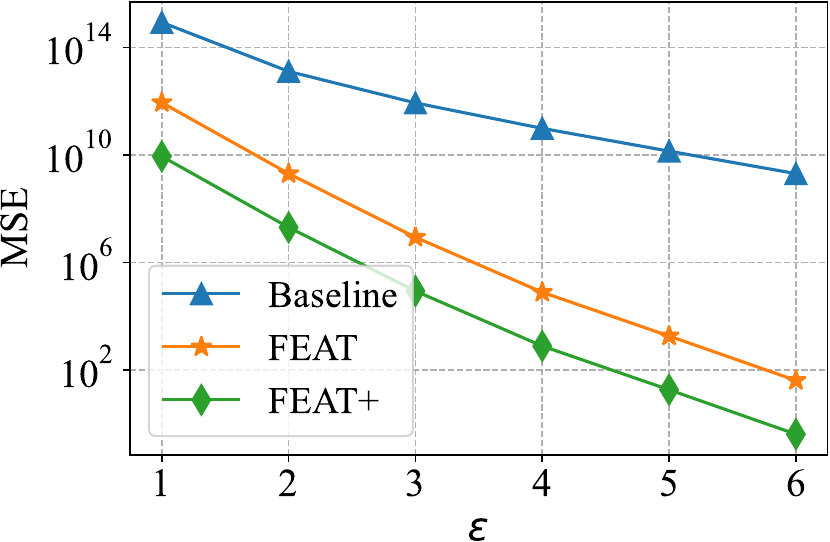}
		\end{minipage}
    \label{fig:triangle mse facebook}
	}%
\subfigure[Wiki]{
		\begin{minipage}[t]{0.49\linewidth}
			\centering
			\includegraphics[width=\linewidth]{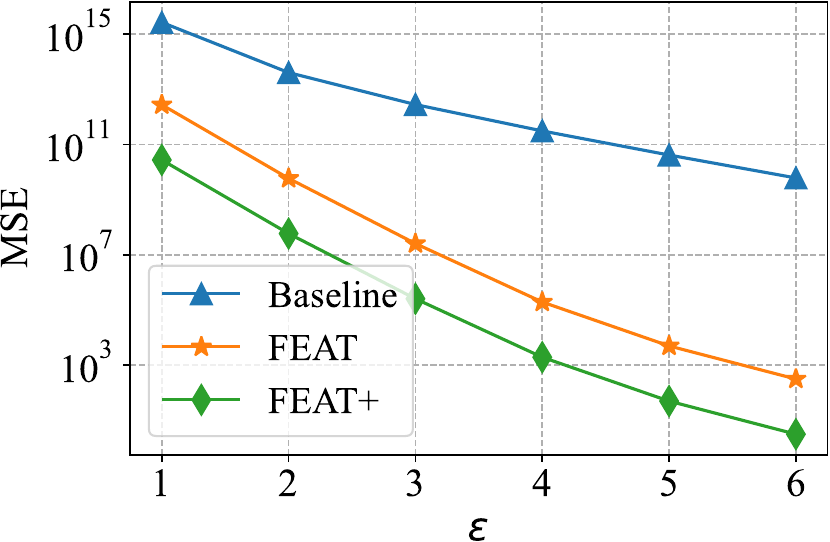}
		\end{minipage}%
	}
       \vspace{-0.4cm}
       \caption{The MSE in triangle counting.}
     \label{fig:triangle mse}
\end{minipage}
 \begin{minipage}[t]{0.49\linewidth}
   \centering
\subfigure[Facebook]{
		\begin{minipage}[t]{0.49\linewidth}
			\centering
			\includegraphics[width=\linewidth]{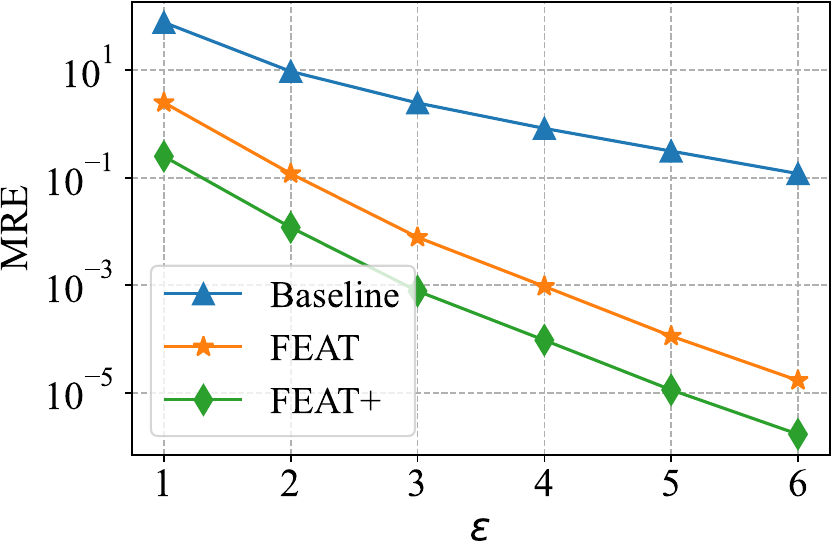}
		\end{minipage}
  \label{fig:triangle mre facebook}
	}%
\subfigure[Wiki]{
		\begin{minipage}[t]{0.49\linewidth}
			\centering
			\includegraphics[width=\linewidth]{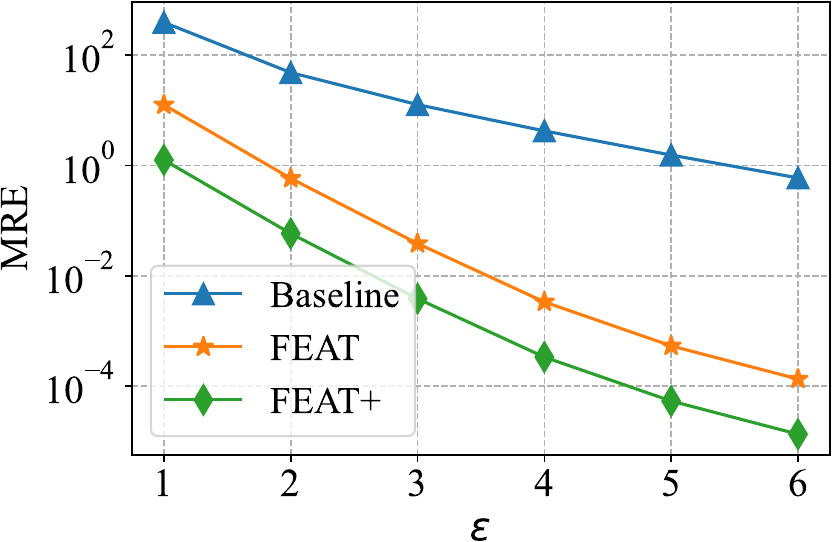}
		\end{minipage}%
  \label{fig:triangle mre wiki}
	}
       \vspace{-0.4cm}
       \caption{The MRE in triangle counting.}
       \label{fig:triangle mre}
\end{minipage}
   \end{center}
        \vspace{-0.3cm}
\end{figure*}

\begin{figure*}[t]
 \begin{center}
     \begin{minipage}[t]{0.49\linewidth}
   \centering
\subfigure[2-Star counting]{
		\begin{minipage}[t]{0.49\linewidth}
			\centering
			\includegraphics[width=\linewidth]{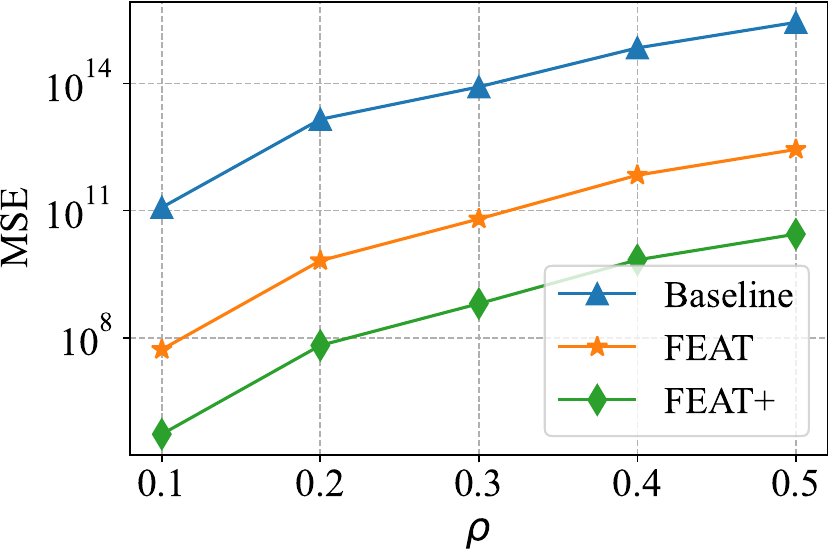}
		\end{minipage}
  \label{fig:sample 2star mse}
	}%
\subfigure[Triangle counting]{
		\begin{minipage}[t]{0.49\linewidth}
			\centering
			\includegraphics[width=\linewidth]{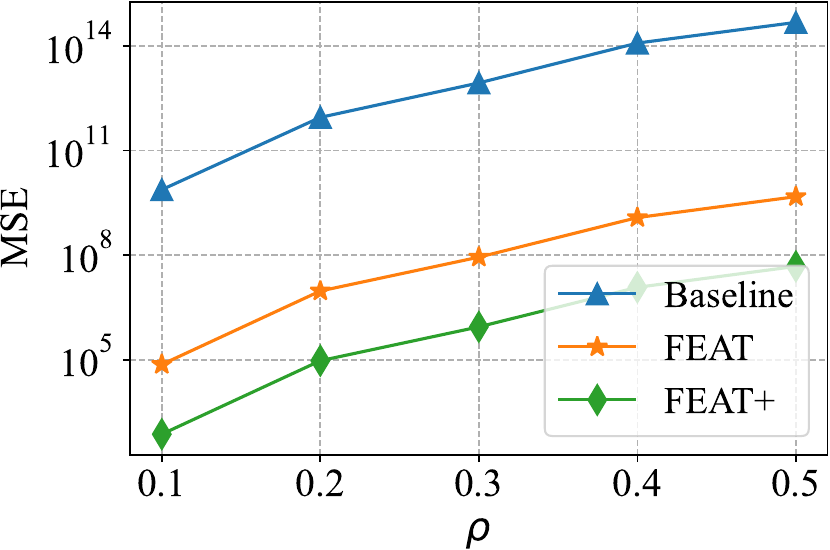}
		\end{minipage}%
    \label{fig:sample triangle mse}
	}
       \vspace{-0.4cm}
       \caption{The MSE with various $\rho$.}
     \label{fig:sample mse}
\end{minipage}
 \begin{minipage}[t]{0.49\linewidth}
   \centering
\subfigure[2-Star counting]{
		\begin{minipage}[t]{0.49\linewidth}
			\centering
			\includegraphics[width=\linewidth]{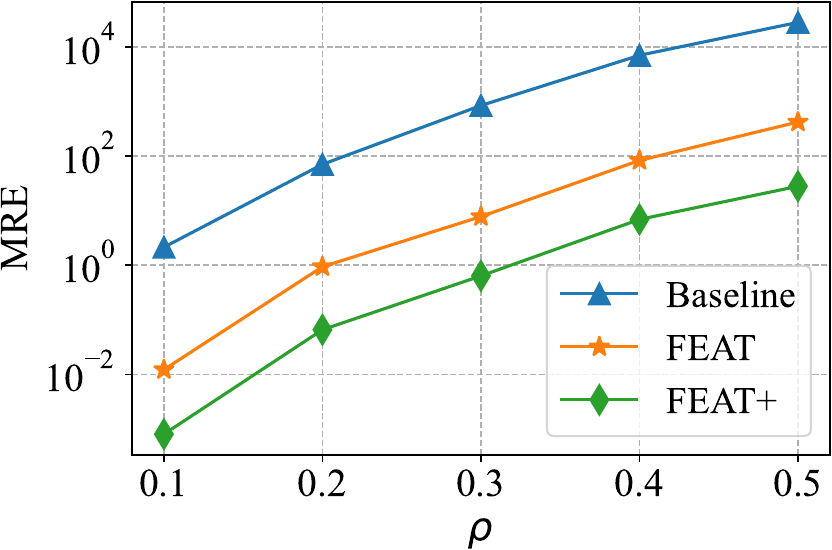}
		\end{minipage}
  \label{fig:sample 2star mre}
	}%
\subfigure[Triangle counting]{
		\begin{minipage}[t]{0.49\linewidth}
			\centering
			\includegraphics[width=\linewidth]{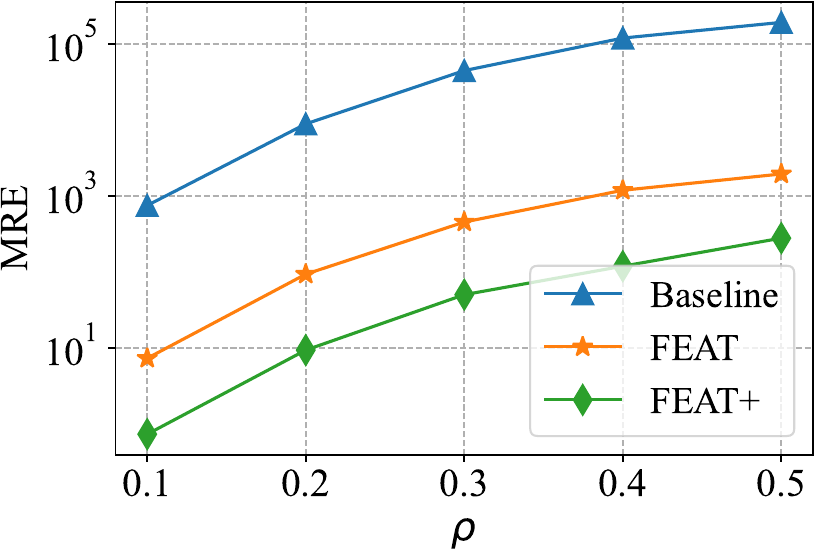}
		\end{minipage}%
  \label{fig:sample triangle re}
	}
       \vspace{-0.4cm}
       \caption{The MRE with various $\rho$.}
       \label{fig:sample mre}
\end{minipage}
   \end{center}
        \vspace{-0.3cm}
\end{figure*}

\begin{figure*}[t]
 \begin{center}
     \begin{minipage}[t]{0.49\linewidth}
   \centering
\subfigure[2-Star counting]{
		\begin{minipage}[t]{0.49\linewidth}
			\centering
			\includegraphics[width=\linewidth]{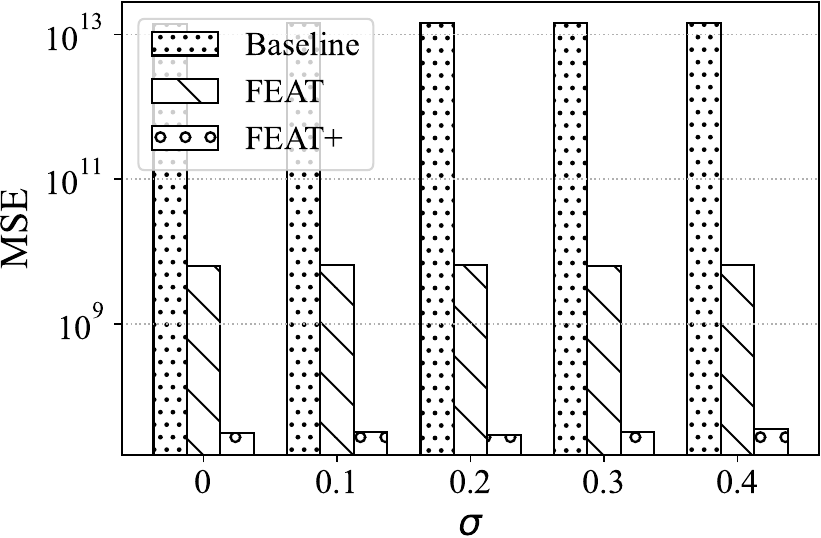}
		\end{minipage}
       \label{fig:overlap 2star mse}
	}%
\subfigure[Triangle counting]{
		\begin{minipage}[t]{0.49\linewidth}
			\centering
			\includegraphics[width=\linewidth]{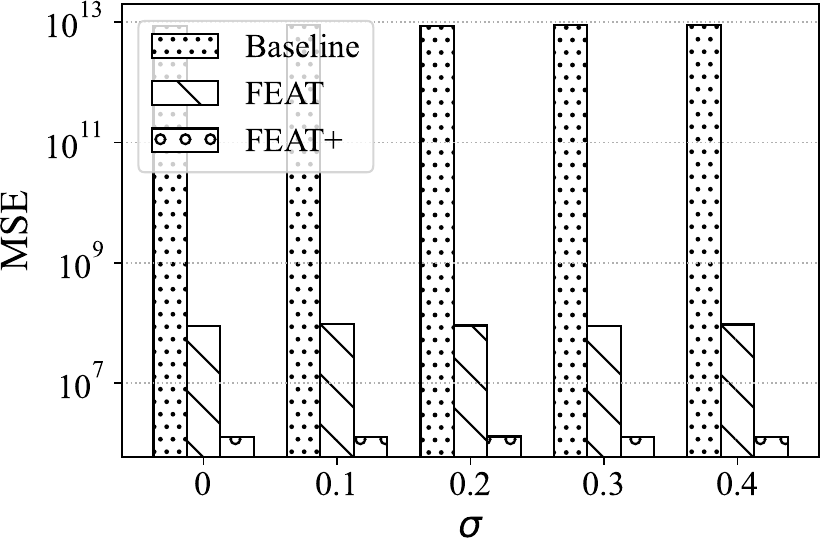}
		\end{minipage}%
       \label{fig:overlap triangle mse}
	}
       \vspace{-0.4cm}
       \caption{The MSE with various $\sigma$.}
     \label{fig:overlap mse}
\end{minipage}
 \begin{minipage}[t]{0.49\linewidth}
   \centering
\subfigure[2-Star counting]{
		\begin{minipage}[t]{0.49\linewidth}
			\centering
			\includegraphics[width=\linewidth]{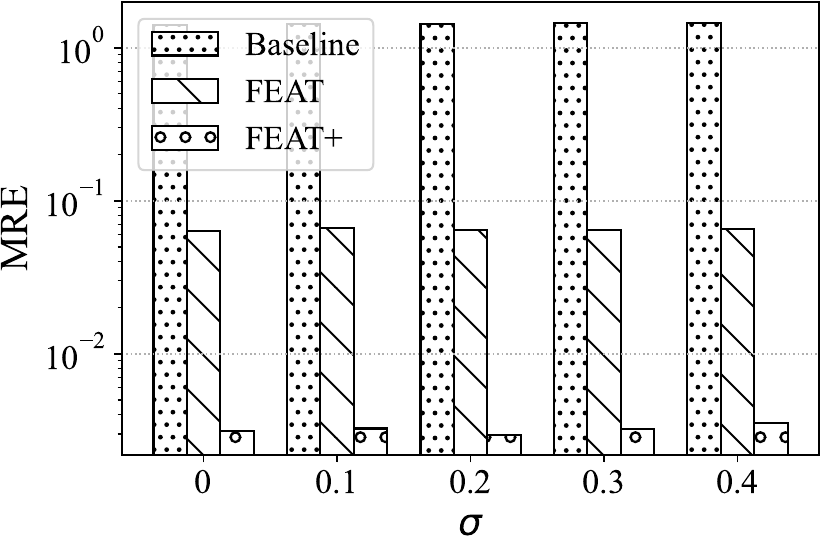}
		\end{minipage}
  \label{fig:overlap 2star mre}
	}%
\subfigure[Triangle counting]{
		\begin{minipage}[t]{0.49\linewidth}
			\centering
			\includegraphics[width=\linewidth]{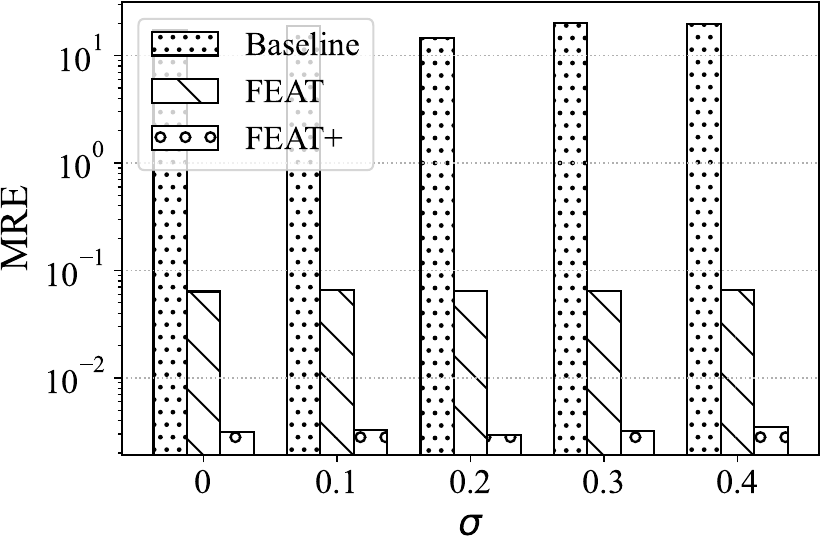}
		\end{minipage}%
  \label{fig:overlap triangle mre}
	}
       \vspace{-0.4cm}
       \caption{The MRE with various $\sigma$.}
       \label{fig:overlap mre}
\end{minipage}
   \end{center}
       \vspace{-0.3cm}
\end{figure*}

\smallskip{}
\noindent\textbf{Evaluation Highlights}:
\begin{itemize}
      \item $\mathsf{FEAT}$ reduces the error of $\mathsf{Baseline}$ by up to an order of 4. $\mathsf{FEAT}$+ outperforms $\mathsf{FEAT}$ by at least an order of 1 (Figures~\ref{fig:2star mse} to \ref{fig:triangle mre}).
      \item $\mathsf{FEAT}$ and $\mathsf{FEAT}$+ significantly outperform $\mathsf{Baseline}$ and $\mathsf{FEAT}$, respectively, under various values of $\rho$ and $\sigma$ (Figures \ref{fig:sample mse} and \ref{fig:overlap mre}).
      \item $\mathsf{FEAT}$ and $\mathsf{FEAT}$+ takes more time than $\mathsf{Baseline}$ by at least an order of 1 (Figure \ref{fig:time}).
\end{itemize}

\subsection{Experimental Setup}
\textbf{Datasets}.
We use two real-world graph datasets from SNAP \cite{snapnets} as follows:
(1) \emph{Facebook}. 
The Facebook graph is collected from survey participants using the Facebook app, which includes 4039 nodes and 88234 edges.
The average degree of the Facebook graph is 21.85 (=$\frac{88234}{4039}$).
(2) \emph{Wiki-Vote}. 
The Wiki-Vote graph contains all the Wikipedia voting data from the inception of Wikipedia till 2008, which includes 7115 nodes and 103689 edges.
The average degree of Wiki-Vote graph is 14.57 (=$\frac{103689}{7115}$).
Thus, Wiki-Vote graph is more sparse than the Facebook graph. 
As explained above, we split a graph randomly into 4 local subgraphs by controlling the sampling rate $\rho$ and overlapping rate $\sigma$.

\smallskip{}
\noindent\textbf{Parameters}.
There are some key parameters that influence the overall accuracy of the $\mathsf{FEAT}$ system. 
(1) \emph{Privacy Budget} $\varepsilon$. 
The privacy budget varies from 1 to 6, and the default is 3.
Note that $\mathsf{FEAT}$+ involves three kind of privacy budgets, namely, $\varepsilon=\varepsilon_1+\varepsilon_2+\varepsilon_3$.
We set $\varepsilon_1=\varepsilon_3=0.45\varepsilon$ and $\varepsilon_2=0.1\varepsilon$.
(2) \emph{Sampling rate} $\rho$.
The sampling rate is the ratio of each local subgraph to the global graph. It is set from 0.1 to 0.5, and the default is 0.3.
(3) \emph{Overlapping Rate} $\sigma$.
The overlapping rate is the ratio of edges shared among multiple local subgraphs to the total number of edges. It varies from 0 to 0.4, and the default is 0.2.

\smallskip{}
\noindent\textbf{Graph Statistics and Metrics}.
For graph analytic tasks, we evaluate two common graph statistics: 2-star counts and triangle counts, as in \cite{ding2021differentially,imola2021locally,imola2022communication,liu2022collecting,imola2022differentially,liu2023cargo}.
For each query, we compare the results $Q$ and $Q^\prime$ from the true graph and noisy graph respectively.
We use two common measures to assess the accuracy of our algorithms: mean squared error (MSE)~\cite{das2004mean} and mean relative error (MRE) \cite{foss2001mre}.
We evaluate the average results over 10 repeated runs.

\subsection{Experimental Results}
\textbf{Utility-Privacy Trade-off (Q1)}.
We first answer Q1 by comparing the utility of our $\mathsf{FEAT}$ and $\mathsf{FEAT}$+ with that of $\mathsf{Baseline}$ when the privacy budget varies from $1$ to $6$.
Figures \ref{fig:2star mse} to \ref{fig:triangle mre} show that the utility loss of all methods decreases with an increase in $\varepsilon$.
Our general framework $\mathsf{FEAT}$ significantly outperforms $\mathsf{Baseline}$ in all cases, and our improved framework $\mathsf{FEAT}$+ can further improve the accuracy of $\mathsf{FEAT}$.

In particular, $\mathsf{FEAT}$ outperforms $\mathsf{Baseline}$ by at least an order of 1.
For instance, for 2-star counting, Figure \ref{fig:2star mse wiki} shows that when $\varepsilon=6$, $\mathsf{FEAT}$ owns an MSE of $5.09\times10^{5}$ while $\mathsf{Baseline}$ gives an MSE of $1.45\times10^{11}$.
Similarly, for triangle counting, Figure \ref{fig:triangle mre facebook} shows that $\mathsf{FEAT}$ gives an MRE of only 9.53$\times10^{-4}$ when $\varepsilon=4$.
In contrast, $\mathsf{Baseline}$ has an MRE of 8.27$\times 10^{-1}$.
This is because $\mathsf{Baseline}$ uses small privacy budgets (i.e., $\varepsilon/m$) to perturb sensitive data, which leads to much noise.
Instead, $\mathsf{FEAT}$ collects noisy graphs using $\varepsilon$ by combining PSU and DP.
The same information is collected only once, and therefore, individual privacy is not leaked.

Another observation is that the error of $\mathsf{FEAT}$+ is smaller than that of $\mathsf{FEAT}$ in all cases.
For example, Figure \ref{fig:2star mre wiki} shows that when $\varepsilon = 6$, $\mathsf{FEAT}$+ only owns an MRE of 3.86$\times 10^{-5}$, whereas $\mathsf{FEAT}$ has an MRE of 2.98$\times 10^{-4}$.
Similarly, Figure \ref{fig:triangle mse facebook} shows that when $\varepsilon = 3$, the MSE of $\mathsf{FEAT}$+ is 4.68$\times 10^{2}$ when $\mathsf{FEAT}$ owns a MSE of 6.25$\times 10^{4}$.
This is mainly because $\mathsf{FEAT}$+ calculates graph statistics by utilizing local true subgraphs.
In contrast, the results of $\mathsf{FEAT}$ are computed based on noisy graphs, which results in much utility loss.
Thus, $\mathsf{FEAT}$ outperforms $\mathsf{Baseline}$ significantly, and $\mathsf{FEAT}$+ improves the utility of $\mathsf{FEAT}$.

\smallskip{}
\noindent\textbf{Parameter Effects (Q2)}.
Next, we evaluate the key parameters that may influence the overall utility of $\mathsf{FEAT}$, i.e., sampling rate $\rho$ and overlapping rate $\sigma$.
$\rho$ determines the size of local subgraphs $G_i, i\in[m]$, and the size of $G_i$ becomes larger as $\rho$ increases.
$\sigma$ determines the number of the same edges that exist in multiple local subgraphs.

Figures \ref{fig:sample mse} and \ref{fig:sample mre} show that in all cases, the MSE and MRE increase with an increase in $\rho$.
This is because graph analytic tasks are executed over a noisy graph $G^\prime$.
The added error becomes larger as the graph size increases.
We can also observe that $\mathsf{FEAT}$ and $\mathsf{FEAT}$+ owns better utility than $\mathsf{Baseline}$ and $\mathsf{FEAT}$, respectively, for all $\rho$.

Figures \ref{fig:overlap mse} and 
\ref{fig:overlap mre} show that $\mathsf{Baseline}$ performs the worst while $\mathsf{FEAT}$+ owns the best utility over all values of the overlapping rate $\sigma$.
Another interesting observation is that the overlapping rate $\sigma$ has little influence on the overall utility.
For instance, Figure~\ref{fig:overlap 2star mse} shows that with the increase in $\sigma$, the MSE of $\mathsf{Baseline}$ increases from 1.014$\times10^{16}$ to 1.0442$\times10^{16}$.
The slight growth is from multiple perturbations of the same edge information.
There are little changes in the results of $\mathsf{FEAT}$ and $\mathsf{FEAT}$+ over various $\sigma$.
This is because the same information is randomized and collected only once.

\begin{figure}[t]
   \centering
\subfigure[Facebook]{
		\begin{minipage}[t]{0.49\linewidth}
			\centering
			\includegraphics[width=\linewidth]{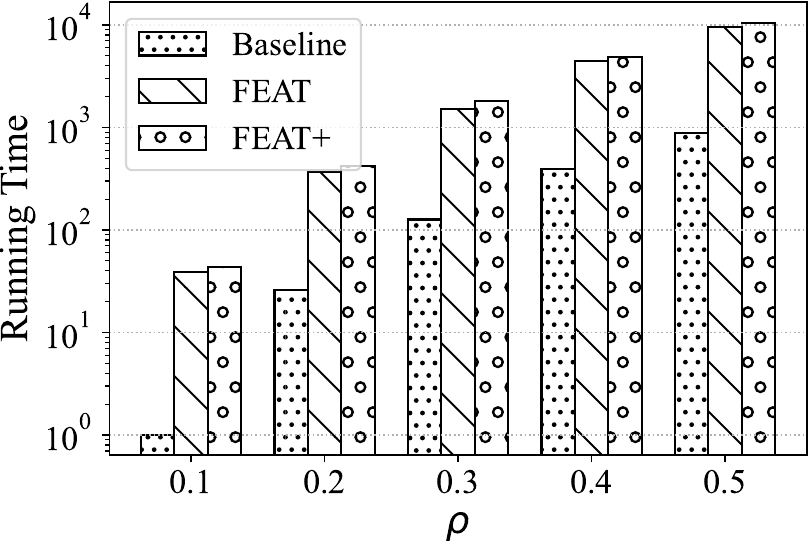}
		\end{minipage}
  \label{fig:time facebook}
	}%
\subfigure[Wiki]{
		\begin{minipage}[t]{0.49\linewidth}
			\centering
			\includegraphics[width=\linewidth]{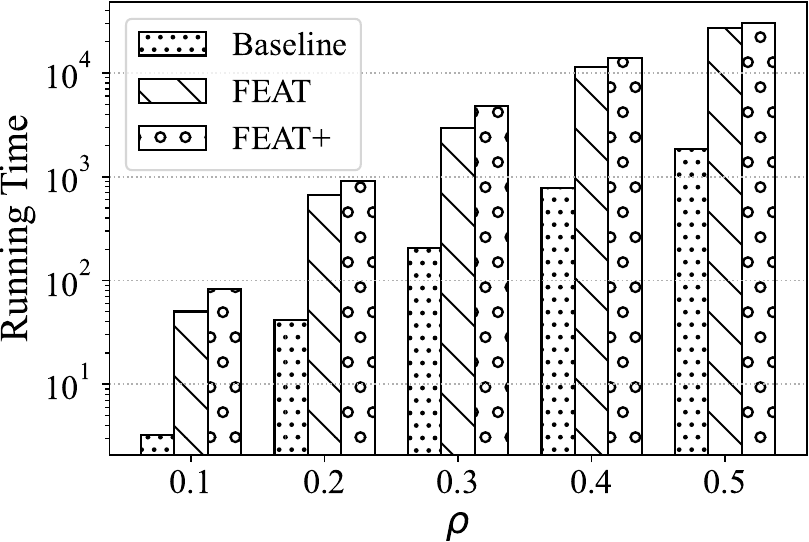}
		\end{minipage}%
     \label{fig:time wiki}
	}
       \vspace{-0.4cm}
       \caption{Running time with various $\rho$.}
        \label{fig:time}
      \vspace{-0.4cm}
\end{figure}

 \smallskip{}
\noindent\textbf{Execution Time (Q3)}.
Finally, we answer the third question by evaluating the running time over graphs with different sizes.
Here, we use different sampling rates $\rho$ to generate multiple graphs with different scales.
Figure \ref{fig:time} presents the running time of $\mathsf{Baseline}$, $\mathsf{FEAT}$, and $\mathsf{FEAT}$+ for various values of  $\rho$.
We can find that the running time increases when the graph scale becomes larger.
This is because when $\rho$ increases, there are more edges collected and computed accordingly.
Another important observation is that the running time of $\mathsf{FEAT}$ is approximately 10$\times$ higher than that of $\mathsf{Baseline}$.
This is because the computation of cryptographic techniques leads to additional time overhead.
$\mathsf{FEAT}$+ takes more time than $\mathsf{FEAT}$ by about 50\%.
This is because additional communication between clients and the server consumes more time.
Thus, utilizing cryptographic tools improves the utility while not leaking sensitive information but at the cost of efficiency.

\section{Related Works}
\label{sec:related works}

\textbf{Federated Analytics}.
The term ``federated analytics'' is first introduced by Google in 2020 \cite{google2020}, which is explored in support of federated learning for Google engineers to measure the quality of federated learning models against real-world data.
Bharadwaj $et$ $al.$ \cite{bharadwaj2022introduction} introduces the notion of federated computation, which is a means of working with private data at a rather large scale.
Wang $et$ $al.$ \cite{wang2021federated} clarify what federated analytics is and its position in literature and then presents the motivation, application, and opportunities of federated analytics.
Elkordy $et$ $al.$ \cite{elkordy2023federated} gives a comprehensive survey about federated analytics. 
Nevertheless, they only focus on tabular data analytics, which totally differs from graph analytics in our work.
Roth  $et$ $al.$ \cite{roth2021mycelium} introduce Mycelium for large-scale distributed graph queries with differential privacy.
Yuan  $et$ $al.$ \cite{yuan2021subgraph} define the notion of graph federation for subgraph matching, where graph data sources are temporarily federated.
However, they assume that all clients are mutually independent, which is different from ours.
Our (informal) previous work~\cite{liu2023federated}\footnote{Note that \cite{liu2023federated} was not published in proceedings in accordance with the policy of the KDD conference, cf. \url{https://fl4data-mining.github.io/calls/}.} introduces the concept of federated graph analytics for the first time. However, it encountered privacy leakage issues due to the potential disclosure of intersection information between two clients.

\smallskip{}
\textbf{Cross-silo Federated Learning}.
There exist several works related to cross-silo federated learning \cite{huang2021personalized,liu2022privacy,zheng2023secure,tang2021incentive,wang2021efficient,lipractical}.
Huang $et$ $al.$\cite{huang2021personalized} propose FedAMP that employs federated attentive message passing to facilitate the collaboration
effectiveness between clients without infringing their
data privacy.
Li $et$ $al.$ \cite{lipractical} propose a practical one-shot federated learning algorithm by using the knowledge transfer technique.
Liu $et$ $al.$ \cite{liu2022privacy} empirically show that MR-MTL is a remarkably strong baseline under silo-specific sample-level DP.
Tang $et$ $al.$ \cite{tang2021incentive} propose an incentive mechanism for cross-silo federated learning, addressing the public goods feature.
Zheng $et$ $al.$ \cite{zheng2023secure} propose a one-server solution based solely on homomorphic encryption and a two-server protocol based on homomorphic encryption and additive
secret sharing, which are designed for contribution
evaluation in cross-silo federated learning.
However, these protocols are designed for machine learning cannot be used for graph analytics.

\smallskip{}
\textbf{Differentially Private Graph Analytics}.
The standard way to calculate graph statistics is through differential privacy (DP) \cite{dwork2014algorithmic,li2016differential}, which is a golden standard in the privacy community. 
However, existing protocols \cite{ding2021differentially,day2016publishing,imola2021locally,imola2022communication,liu2022collecting,imola2022differentially,liu2022crypto} are designed for different scenarios and not applicable to federated graph scenarios.
To be specific, central protocols \cite{ding2021differentially,day2016publishing} rely on a trusted server to collect the entire graph information from local users and then release accurate graph statistics privately.
Nevertheless, a centrally trusted server is amenable to privacy issues in practical such as data leaks and breaches \cite{neto2021developing,gibson2021vulnerability}.
Instead, local protocols \cite{imola2021locally,imola2022communication} remove the assumption of a trusted server, and each user directly perturbs local sensitive data.
However, they cannot protect the user-client membership and faces the edge privacy issues due to the overlaps.
Extended local view (ELV)-based protocols \cite{liu2022collecting,sun2019analyzing} consider an extension of local scenarios, where each client can see not only her 1-hop path but also her 2-hop path.
Similar to local mechanisms, ELV-based protocols fail to protect the user-client membership.
Additionally, decentralized differential privacy in \cite{sun2019analyzing} can protect edge privacy in ELV but fails to do it in federated settings since overlaps of ELV are different from those of federated graph analytics.
In a nutshell, this paper is the first work to formulate the federated graph analytics (FGA) to our best of knowledge.
Our proposed $\mathsf{FEAT}$ is a general framework for various common graph analytics.

\section{Conclusion}
\label{sec:conclusion}
We made the first attempt to study federated graph analytics with privacy guarantee. 
We showed unique challenges in federated graph
analytics, namely, utility issue from the limited view and privacy
issue due to overlaps.
To alleviate them, we proposed a general federated
graph analytic framework $\mathsf{FEAT}$, based on our proposed differentially private set union protocol.
Furthermore, we observed that it calculates graph statistics over a noisy global graph without considering true local subgraphs, and there is still room for improving the overall utility.
To address this issue, we introduced an improved framework $\mathsf{FEAT}$+ by combining a noisy global graph with true local subgraphs.
Comprehensive experiments verify that our proposed methods significantly outperform the baseline approaches in various graph analytics.

\bibliographystyle{IEEEtran}
\bibliography{reference}




\newpage

 

\end{document}